\documentclass[article,10pt,nofootinbib,notitlepage,twocolumn]{revtex4-1}
\usepackage{graphicx}
\usepackage{amsmath}
\usepackage{amssymb}
\usepackage{dcolumn}
\usepackage{bm}
\usepackage{color}
\usepackage{dsfont}
\usepackage{feynmp}
\usepackage{marvosym}
\usepackage{phaistos}

\newcommand \beq{\begin{eqnarray}}
\newcommand \eeq{\end{eqnarray}}

\begin{document}
\unitlength=1mm
\allowdisplaybreaks

\title{Center-symmetric Landau gauge: Further signatures of confinement}

\author{Duifje Maria van Egmond}
\affiliation{Centre de Physique Th\'eorique, CNRS, Ecole polytechnique, IP Paris, F-91128 Palaiseau, France.}

\author{Urko Reinosa}
\affiliation{Centre de Physique Th\'eorique, CNRS, Ecole polytechnique, IP Paris, F-91128 Palaiseau, France.}

\date{\today}

\begin{abstract}
In a recent article \cite{vanEgmond:2022nuo}, we have identified new signatures for the Yang-Mills deconfinement transition, based on the finite-temperature {\it longitudinal} or {\it (chromo-)electric} gluon propagator as computed in the center-symmetric Landau gauge. Here, we generalize these considerations into a systematic study of the center symmetry identities obeyed by the correlation functions in this gauge. Any violation of these constraints signals the breaking of center symmetry and can thus serve as a probe for the deconfinement transition.
\end{abstract}

\maketitle

\section{Introduction}
Functional methods are by now a well developed corpus of approaches in the framework of non-abelian gauge theories \cite{vonSmekal:1997ohs, Alkofer:2000wg, Zwanziger:2001kw, Fischer:2003rp, Bloch:2003yu, Aguilar:2004sw, Boucaud:2006if, Aguilar:2007ie, Aguilar:2008xm, Boucaud:2008ky, Fischer:2008uz, Rodriguez-Quintero:2010qad,Wetterich:1992yh, Berges:2000ew, Pawlowski:2003hq, Fischer:2004uk,Pawlowski:2005xe,Cyrol:2016tym,Dupuis:2020fhh} that can bring valuable complementary information in situations where Monte-Carlo lattice simulations are the least efficient. One limitation of functional methods, as compared to the lattice, is, however, that the primary quantities they give access to are gauge-dependent correlation functions. Although observables can be reconstructed from the correlation functions in principle, this strongly rests on the accuracy at which the latter are computed and on their particular relation to the observables under consideration.

A natural question that emerges is then whether it could be possible to extract relevant physical information directly from the correlation functions themselves, an idea that can be further elaborated in at least two distinct directions. The most obvious one is to try to identify gauge-independent features of the correlation functions \cite{Kobes:1990dc}, which are then more prone to encapsulate observable information. A different strategy is based on the idea that certain physical questions could be addressed directly from gauge-dependent features of the correlation functions, in certain, well chosen gauges. A paradigmatic example is the question of symmetries and their breaking, which usually underlies the phase structure of the system under consideration. 

In particular, in recent years, both lattice simulations \cite{Cucchieri:2010lgu,Fischer:2010fx,Maas:2011ez,Mendes:2015jea,Aouane:2011fv,Silva:2013maa} as well as various analytical studies \cite{Fister:2011uw, Fischer:2012vc, Fukushima:2013xsa, Huber:2012kd, Quandt:2015aaa} have searched for signatures of the Yang-Mills deconfinement phase transition within the Landau gauge gluon propagator. However, even though some change of behavior is seen in the vicinity of the transition and some quantities extracted from the propagator seem to behave as order parameters, no clear connection has been established with the breaking of center symmetry.\footnote{A very interesting step in this direction has been taken in Ref.~\cite{Silva:2016onh} where a connection between center-symmetry breaking and a ``sectorized" gluon propagator (defined by classifying gluon field configurations according to the domain of the complex plane in which the Polyakov lies) has been pointed out. To date, unfortunately, this is not an easy quantity to evaluate in the continuum.} Moreover, in the particular case of the SU($2$) gauge group, where the transition is known to be second order, the Landau gauge gluon propagator does not seem to become critical.

In order to explain these observations, in a series of works \cite{vanEgmond:2021jyx,vanEgmond:2022nuo,vanEgmond:2023lnw}, we have put forward the idea that the Landau gauge might not be the most appropriate gauge to analyze the deconfinement transition, in particular when it comes to identifying signatures of symmetry breaking at the level the correlation functions. One can actually understand this on very general grounds since, as we argue below, the gauge-fixed action associated to a chosen gauge fixing does not necessarily reflect the symmetries of the problem. Even though this feature has no influence on the way the symmetry constrains the observables,\footnote{At least at an exact level of treatment, see the general discussion in Refs.~\cite{Reinosa:2020mnx,vanEgmond:2023lnw}.} it can have a strong impact on whether and how the symmetries manifest themselves at the level of the correlation functions.

To be more specific, consider a gauge theory defined by some non-gauge-fixed action $S[A]$ and consider a physical symmetry $\smash{A\to A'}$ such that $\smash{S[A']=S[A]}$. When switching to a gauge-fixed formulation, that same physical transformation does not need to be a symmetry of the gauge-fixed action $S_{\rm gf}[A]$. In fact, one could more generally have\footnote{For a thorough discussion of these questions, see Refs.~\cite{Reinosa:2020mnx,vanEgmond:2023lnw}.}
\beq
S_{\rm gf}[A']=S_{\rm gf'}[A]\,,\label{eq:rel}
\eeq
with ${\rm gf}'$ a gauge fixing that can differ from the original one ${\rm gf}$. Indeed, although not representing a symmetry of the action in a given gauge, the identity (\ref{eq:rel}) is actually sufficient for the physical symmetry to be manifest at the level of the observables $\langle{\cal O}[A]\rangle_{\rm gf}$. For instance, assuming a linear transformation 
\beq
{\cal O}[A']=L{\cal O}[A]\,,\label{eq:lin}
\eeq 
and because the observables do not depend on the gauge fixing, we can write the following chain of identities:\footnote{The first equality is just the fact that $A$ is a dummy integration variable under the functional integral, the second equality uses both (\ref{eq:rel}) and (\ref{eq:lin}), and the third equality uses the gauge-fixing independence of the observables.}
\beq
\langle{\cal O}[A]\rangle_{\rm gf}=\langle{\cal O}[A']\rangle_{\rm gf}=L\langle{\cal O}[A]\rangle_{\rm gf'}=L\langle{\cal O}[A]\rangle_{\rm gf}\,.
\eeq
This gives a symmetry constraint for the considered observable in any chosen gauge and can then be used as a probe for the breaking of the symmetry under consideration. A well known example of such type of observables is the Polyakov loop that probes the center-symmetry in pure Yang-Mills theories.

If we consider instead a correlation function $\langle{\cal C}[A]\rangle_{\rm gf}$, and assuming again a linear transformation ${\cal C}[A']=L{\cal C}[A]$, the previous chain of identities stops one step earlier because correlation functions depend explicitly on the gauge fixing:
\beq
\langle{\cal C}[A]\rangle_{\rm gf}=\langle{\cal C}[A']\rangle_{\rm gf}=L\langle{\cal C}[A]\rangle_{\rm gf'}\,.\label{eq:chain}
\eeq
In this case, one obtains a relation between the correlation functions in the two different gauges ${\rm gf}$ and ${\rm gf}'$ but certainly not a constraint on the correlation functions of a given gauge. For this reason, the correlation functions cannot in general be used as probes for the breaking of physical symmetries.

There is one important exception, however, corresponding to the case where the chosen gauge fixing is invariant under the considered physical transformation, that is $\smash{{\rm gf}'={\rm gf}}$. In this case, the chain of identities (\ref{eq:chain}) can be continued one step further, just as for the case of observables (but not for the same reason), and one obtains symmetry constraints for the correlation functions themselves, that can be used as order parameters for the symmetry at hand. 

In this work, we consider one particular example of such symmetry invariant gauge fixings, the recently introduced center-symmetric Landau gauges \cite{vanEgmond:2021jyx} which are invariant under the center symmetry of pure SU(N) Yang-Mills theories at finite temperature \cite{Gavai:1982er,Gavai:1983av,Celik:1983wz,Svetitsky:1985ye,Pisarski:2002ji,Greensite:2011zz} and which are thus adapted to the study of the deconfinement transition from the correlation functions. Some of these aspects were analyzed in Ref.~\cite{vanEgmond:2022nuo} using specific Lorentz/color projections of the gluon two-point function. The present work makes the discussion more general by extending it to any correlation function and any Lorentz/color projection.

The article is organized as follows. In the next section, we define the notion of center-symmetric gauge field backgrounds and the associated center-symmetric Landau gauges. We also particularize to backgrounds (and thus, gauges) that are, in addition, charge conjugation invariant, and also invariant under particular color rotations. Section \ref{sec:color} analyzes the constraints on correlation functions associated to color invariance. The discussion of charge conjugation and center symmetry is more subtle because one needs to pay attention to the fact that physical symmetries act on gauge fields modulo genuine gauge transformations. One convenient way to handle this aspect of the theory is through the notion of Weyl transformations and Weyl chambers which we discuss in Sec.~\ref{sec:weyl}. Sections.~\ref{sec:C_constraints}, \ref{sec:Z2_constraints} and \ref{sec:Z3_constraints} are then devoted to a systematic analysis of the constraints on the correlation functions that derive from charge conjugation and center symmetry for $\smash{N=2}$ and $\smash{N=3}$, as well as the identification of new order parameters for center symmetry. Additional details are gathered in the Appendixes.

\section{Center-symmetric Landau gauges}

In what follows, we consider SU(N) Euclidean Yang-Mills theories within the framework of {\it background Landau gauges} \cite{Abbott:1980hw,Abbott:1981ke,Braun:2007bx, Braun:2010cy}. The latter actually refers to a family of gauges parametrized by a background gauge field configuration $\bar A$ that, in a sense, plays the role of an infinite collection of gauge-fixing parameters. The gauge condition is
\beq
\bar D_\mu(A_\mu^a-\bar A_\mu^a)=0\,,\label{eq:cond}
\eeq
where
\beq
\bar D_\mu\varphi^a\equiv\partial_\mu\varphi^a+gf^{abc}\bar A_\mu^b\varphi^c
\eeq
stands for the adjoint covariant derivative in the presence of the background. 

A given background defines a particular choice of gauge within the class of background Landau gauges. For instance, when the background is taken equal to zero, one retrieves the standard Landau gauge $\smash{\partial_\mu A_\mu^a=0}$. Here, we are interested in the subclass of {\it center-symmetric Landau gauges} \cite{vanEgmond:2021jyx} obtained by choosing, instead, center-symmetric backgrounds which we now define in more detail.

\subsection{Center-symmetric backgrounds}
A {\it center-symmetric background} $\bar A_c$ is defined by the condition
\beq
\forall U\in {\cal G},\, \quad \exists U_0\in {\cal G}_0\,, \quad \bar A^{U_0U}_c=\bar A_c\,.\label{eq:def}
\eeq
Here, ${\cal G}$ denotes the group of gauged SU(N) matrices $U(\tau,\vec{x})$ obeying the particular boundary conditions\footnote{As usual, $\smash{\beta\equiv 1/T}$ denotes the inverse temperature and corresponds to the extent of the Euclidean time interval over which the fields are defined.}
\beq
U(\tau+\beta,\vec{x})=e^{i2\pi k/N}U(\tau,\vec{x})\,,\label{eq:boundary}
\eeq
with $\smash{k=0,1,\dots,N-1}$, while ${\cal G}_0$ is the subgroup of ${\cal G}$ corresponding to $k=0$. 

Any $\smash{U\in{\cal G}}$ acts on the gauge field as
\beq
A_\mu^U\equiv UA_\mu U^\dagger+\frac{i}{g}U\partial_\mu U^\dagger\,,
\eeq
where we have defined $\smash{A_\mu\equiv A_\mu^a t^a}$. It should be stressed, however, that only those $\smash{U_0\in{\cal G}_0}$ correspond to genuine gauge transformations, that is transformations that do not alter the state of the system. In contrast, any $\smash{U\in {\cal G}}$ with $k\neq 0$ transforms at least one observable, the Polyakov loop \cite{Polyakov:1978vu},\footnote{The Polyakov loop is directly related to the free-energy $F_q$ of a static quark in a thermal bath of gluons, $\smash{\ell\sim e^{-\beta F_q}}$. Under the action of $U\in {\cal G}$, it gets multiplied by a phase factor $e^{i2\pi k/N}$.}
\beq
\ell\equiv\frac{1}{N}{\rm tr}\left\langle P\exp\left\{i\int_0^\beta d\tau\,A_0(\tau,\vec{x})\right\}\right\rangle,
\eeq
and should therefore be considered as a physical transformation. 

Actually, because $U$ and $U_0U$ act on the Polyakov loop in the same way, these physical transformations are defined only modulo multiplication by elements of ${\cal G}_0$. This is of course in correspondence with the fact that two gauge field configurations $A$ and $A^{U_0}$ connected by an element of ${\cal G}_0$ should be interpreted as two equivalent representations of the same physical state. In turn, this explains the particular definition of center-invariant configurations given in Eq.~(\ref{eq:def}).\footnote{These considerations apply in fact to any physical symmetry, see the discussion in Refs.~\cite{Reinosa:2020mnx,vanEgmond:2023lnw}.} This definition can actually be replaced by a simpler condition, namely
\beq
\exists U_c\in {\cal U}_1\,, \quad \bar A_c^{U_c}=\bar A_c\,,\label{eq:def2}
\eeq
where ${\cal U}_1$ denotes the set (not a group) of gauged SU(N) matrices that fulfil Eq.~(\ref{eq:boundary}) with $\smash{k=1}$. We shall stick to this simpler formulation in what follows. 

It should also be mentioned that a given center-symmetric background can obey additional symmetries. It can happen for instance that it is invariant under charge conjugation in the following sense:
\beq
\exists U_0\in {\cal G}_0\,, \quad (-\bar A_c^{\rm t})^{U_0}=\bar A_c\,,\label{eq:C}
\eeq
where $X^{\rm t}$ denotes the transposed of $X$. Finally, the background could be invariant under certain elements of ${\cal G}_0$ (in general global transformations):
\beq
\exists U_0\in {\cal G}_0\,, \quad \bar A_c^{U_0}=\bar A_c\,.\label{eq:color}
\eeq
 The center-symmetric backgrounds that we consider below obey such additional symmetries which we also exploit.

\subsection{Symmetry constraints}
The main interest of center-symmetric backgrounds and center-symmetric Landau gauges is that the gauge-fixed action is invariant under center transformations. By this, we mean that
\beq
\exists U_c\in {\cal U}_1\,, \quad S_{\bar A_c}[A]=S_{\bar A_c}[A^{U_c}]\,.\label{eq:sym}
\eeq
This is to be contrasted to what happens for a gauge choice corresponding to an arbitrary background $\bar A$. In this case, one has instead $\smash{S_{\bar A}[A]=S_{\bar A^U}[A^U]}$, with $\smash{\bar A^U\neq \bar A}$ for any $\smash{U\in {\cal U}_1}$. The latter identity connects the gauge-fixed actions in two different gauges, corresponding respectively to $\bar A$ and $\bar A^U$, and is thus quite different from Eq.~(\ref{eq:sym}) which is a symmetry identity within a single gauge, corresponding to the choice $\smash{\bar A=\bar A_c}$. 

The symmetry identity (\ref{eq:sym}) implies constraints on the correlation functions as computed in the center-symmetric Landau gauge. Take first the one-point function $\langle A\rangle_{\bar A_c}$. If the symmetry is not broken, then we must have
\beq
\langle A\rangle_{\bar A_c}=\langle A^{U_c}\rangle_{\bar A_c}=\langle A\rangle_{\bar A_c}^{U_c}\,.
\eeq
This means that the one-point function should also correspond to a center-symmetric configuration. Any departure from this expectation signals the breaking of center symmetry.

Next, if we consider a connected correlation function
\beq
& & \langle A_{\mu_1}^{a_1}(x_1)\cdots A_{\mu_n}^{a_n}(x_n)\rangle^{\rm connected}_{\bar A_c}\nonumber\\
& & \hspace{1.5cm}=\langle \delta A_{\mu_1}^{a_1}(x_1)\cdots \delta A_{\mu_n}^{a_n}(x_n)\rangle_{\bar A_c}\,,\label{eq:cntd}
\eeq
with $\delta A\equiv A-\langle A\rangle_{\bar A_c}$ transforming as
\beq
(\delta A^{U_c})_\mu^a(x) & = & (U_c(x)\delta A_\mu(x) U^\dagger_c(x))^a\nonumber\\
& \equiv & {\cal U}_c^{ab}(x)\delta A_\mu^b(x)\,,
\eeq
we have, when the symmetry is not broken,
\beq
& & \langle A_{\mu_1}^{a_1}(x_1)\cdots A_{\mu_n}^{a_n}(x_n)\rangle^{\rm connected}_{\bar A_c}\nonumber\\
& & \hspace{1.0cm}=\,{\cal U}_c^{a_1b_1}(x_1)\cdots {\cal U}_c^{a_nb_n}(x_n)\nonumber\\
& & \hspace{1.5cm}\times\,\langle A_{\mu_1}^{b_1}(x_1)\cdots A_{\mu_n}^{b_n}(x_n)\rangle^{\rm connected}_{\bar A_c}\,.\label{eq:constraints}
\eeq
It is easily seen that the vertex functions obey similar identities.\footnote{These identities can be obtained more directly by first writing a symmetry constraint on the generating functional for connected correlation functions, and, then, deducing a similar constraint on the associated effective action, see Refs.~\cite{Reinosa:2020mnx,vanEgmond:2023lnw} for details.} These constraints can serve as probes of deconfinement since the violation of any of these identities signals the breaking of center symmetry. 

We stress that the converse is not true as some of these identities could be further protected by other (unbroken) symmetries even when center symmetry breaks. This in particular the case when (\ref{eq:C}) or (\ref{eq:color}) apply. From these equations, one can indeed derive similar constraints as (\ref{eq:constraints}), provided one replaces ${\cal U}_c$ with the appropriate ${\cal U}_C$ or ${\cal U}_0$. We will see below that some of the constraints derived from charge conjugation coincide with some of the constraints derived from center symmetry. Since charge conjugation is not expected to break spontaneously, these particular constraints cannot be used as probes of the breaking of center symmetry.

A remark on notation is now in order. From now on, we shall omit the label ``connected" when writing correlation functions, and neither should we use the label $\bar A_c$. It is implicitly assumed that, with the exception of the one-point function, the notation $\langle A_{\mu_1}^{a_1}(x_1)\cdots A_{\mu_n}^{a_n}(x_n)\rangle$ refers either to a connected correlation function or to a vertex function, computed in the center-symmetric Landau gauge. Moreover, we shall use a condensed notation such that, unless specifically stated, both Lorentz indices and position arguments are combined into one single index.

\subsection{Constant, temporal and diagonal backgrounds}
In practice, one does not need to determine all possible backgrounds complying with Eq.~(\ref{eq:def2}) but it is enough to find just one. Particularly simple examples are obtained by first restricting to backgrounds of the form
\beq
\bar A_\mu(x)=\delta_{\mu 0}\frac{T}{g} \bar r^j t^j\,,\label{eq:bg}
\eeq
where the $t^j$'s provide a maximal set of commuting generators of the algebra. One then looks for specific values $\bar r_c$ of $\bar r$ that correspond to center-symmetric backgrounds. A convenient way to find these particular values is through the use of Weyl transformations and Weyl chambers, see below for further details, as well as the discussion in Ref.~\cite{Reinosa:2020mnx}. 

In the SU($2$) case, one possible choice is
\beq
\bar r_c=\bar r^3_c=\pi\,,\label{eq:pi}
\eeq 
with $\smash{t^j=\sigma^3/2}$, whereas in the SU($3$) case, one can take
\beq
\bar r_c=(\bar r^3_c,\bar r^8_c)=\left(\frac{4\pi}{3},0\right),\label{eq:4pio3}
\eeq 
with $t^j\in\{\lambda_3/2,\lambda_8/2\}$. We shall restrict to these choices in what follows. We will see below that not only do they fulfil Eq.~(\ref{eq:def2}), but they also comply with Eqs.~(\ref{eq:C}) and (\ref{eq:color}).

\section{Color constraints}\label{sec:color}

The presence of a background makes the color structure of the various correlation functions more intricate than in the Landau gauge. For backgrounds of the form (\ref{eq:bg}), the color structure remains simple, however, due to the fact that the background, and thus the gauge-fixed action, is invariant under global color rotations of the form $\smash{U_\theta=e^{i\theta^jt^j}}$. 

This is just a particular example of Eq.~(\ref{eq:color}), with similar consequences on the correlation functions as the constraints (\ref{eq:constraints}) provided one replaces ${\cal U}_c$ with the corresponding ${\cal U}_\theta$. To make the most of this symmetry, and in fact of the other symmetries as well, it will be convenient to work within a Cartan-Weyl basis $\{t^\kappa\}$ whose definition we now recall.

\subsection{Cartan-Weyl bases}

By construction, the generators $t^\kappa$ of a Cartan-Weyl basis simultaneously diagonalize the adjoint action of the $t^j$'s:
\beq
[t^j,t^\kappa]=\kappa^j t^\kappa\,.\label{eq:canonical}
\eeq
It can be helpful to recall these relations using a quantum mechanical language: the labels $\kappa$ are real-valued, $(N-1)$-dimensional vectors that collect the ``quantum (eigen)numbers'' $\kappa^j$ that a given ``(eigen)state'' $t^\kappa$ acquires under the action of the various charges $[t^j,\,\,\,]$. In more technical terms, the vectors $\kappa$ are the {\it adjoint weights} of the algebra.

The adjoint weights can be of two types. If they are non-zero, they are called {\it roots} and are represented using the first letters of the Greek alphabet, $\kappa=\alpha,\,\beta$, \dots The roots are non-degenerate, meaning that there is only one eigenstate $t^\alpha$ associated to a given root $\alpha$. It is also generally true that, if $\alpha$ is a root, then $-\alpha$ is a root as well. Aside from the roots, there is also a vanishing adjoint weight. It is degenerate\footnote{The only exception is the SU($2$) case.} because any $t^{j}$ is an eigenstate with vanishing charges. One can again write these states as $t^\kappa$ provided one sets $\smash{\kappa=0^{(j)}}$. This notation should be understood as representing multiple copies of the null vector, needed to distinguish the various degenerate zero-charge states $\smash{t^{0^{(j)}}=t^j}$. Of course, this label should be interpreted as nothing else but the null vector when appearing in algebraic expressions (that is anytime it is not used as a label).

In what follows, we refer to $\smash{\kappa=0^{(j)}}$ and $\smash{\kappa=\alpha}$ as the {\it neutral} and {\it charged} modes respectively. In the SU($2$) case, we have one neutral mode $0^{(3)}$ and two charged modes $\smash{\alpha_{12}=1}$ and $\smash{\alpha_{21}\equiv -\alpha_{12}}$. In the SU($3$) case, we have two neutral modes, $0^{(3)}$ and $0^{(8)}$, as well as six charged modes $\smash{\alpha_{12}=(1,0)}$, $\smash{\alpha_{23}=(-1/2,\sqrt{3}/2)}$, $\smash{\alpha_{31}=(-1/2,-\sqrt{3}/2)}$, $\smash{\alpha_{21}\equiv-\alpha_{12}}$, $\smash{\alpha_{32}\equiv-\alpha_{23}}$ and $\smash{\alpha_{13}\equiv-\alpha_{31}}$. We note that the roots are all of norm $\smash{\alpha^2=1}$, a property which extends to the SU(N) case, see App.~\ref{app:n}. The notation $\alpha_{kl}$ will be explained further below.

\subsection{Constraints}
From Eq.~(\ref{eq:canonical}), it is easily deduced that the adjoint action of $U_\theta$ on the generators $t^\kappa$ of a Cartan-Weyl basis reads
\beq
U_\theta t^\kappa U^\dagger_\theta=e^{i\theta^j[t^j,\,\,\,]}t^\kappa=e^{i\theta\cdot \kappa}t^\kappa\,,\label{eq:Ut}
\eeq
with $\smash{X\cdot Y\equiv X^jY^j}$. It follows that, within a Cartan-Weyl basis, ${\cal U}_\theta^{\kappa\lambda}=e^{i\theta\cdot \kappa}\delta^{\kappa\lambda}$ and the correlation functions are invariant under the transformation
\beq
A_\mu^\kappa\to {\cal U}^{\kappa\lambda}A_\mu^\lambda=e^{i\theta\cdot\kappa}A_\mu^\kappa\,.
\eeq
The constraints on the correlation/vertex functions take then the form
\beq
\langle A_{\mu_1}^{\kappa_1}\cdots A_{\mu_n}^{\kappa_n}\rangle=e^{i\theta\cdot(\kappa_1+\cdots+\kappa_n)}\langle A_{\mu_1}^{\kappa_1}\cdots A_{\mu_n}^{\kappa_n}\rangle\,.\label{eq:color_constraint}
\eeq
This implies that the non-vanishing correlators are necessarily such that color is conserved in the sense $\smash{\kappa_1+\cdots+\kappa_n=0}$. 

\begin{center}
\begin{figure}[t]
\includegraphics[height=0.25\textheight]{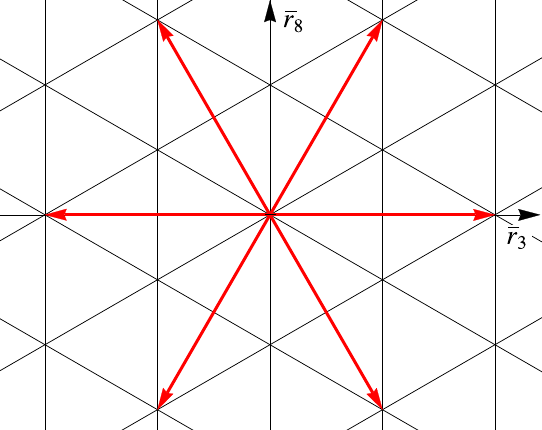}
\caption{SU($3$) Weyl chambers in the $(\bar r_3,\bar r_8)$-plane and their relation to the Weyl transformations and the roots. The red vectors represent the roots multiplied by $4\pi$. The elements of $\tilde {\cal G}_0$ are generated by translations along these vectors and reflections with respect to lines orthogonal to these vectors that go through the origin. Equivalently, they are generated by all possible reflections with respect to lines orthogonal to the roots and displaced by any multiple of $2\pi$ times the corresponding root. The corresponding symmetry axes define a paving of the $(\bar r_3,\bar r_8)$-plane into physically equivalent regions, known as Weyl chambers.}
\label{fig:weyl}
\end{figure}
\end{center}

In particular, the non-vanishing components of the two-point function ${\cal G}_{\mu\nu}^{\kappa\lambda}\equiv\langle A_\mu^\kappa A_\nu^\lambda\rangle$ are necessarily such that $\smash{\kappa+\lambda=0}$. In the SU($2$) case, it immediately follows that $\smash{{\cal G}_{\mu\nu}^{\kappa\lambda}={\cal G}_{\mu\nu}^{\lambda}\delta^{\kappa(-\lambda)}}$. In the SU($3$) case, we cannot conclude this as yet, however, because there are two neutral modes and we could have $\kappa+\lambda=0$ with $\kappa=0^{(3)}$ and $\lambda=0^{(8)}$. What can be said without any further assumption is that 
\beq
{\cal G}_{\mu\nu}^{\alpha 0^{(j)}}={\cal G}_{\mu\nu}^{0^{(j)}\alpha}=0\,,\label{eq:c1}
\eeq 
and 
\beq
{\cal G}_{\mu\nu}^{\alpha\beta}={\cal G}_{\mu\nu}^\beta\delta^{\alpha(-\beta)}\,,\label{eq:c2}
\eeq
but the structure of ${\cal G}_{\mu\nu}^{0^{(j)}0^{(j')}}$ needs still to be investigated.

Let us now study, in a similar way, the constraints deriving from charge conjugation and center invariance. Their study is slightly more delicate because the corresponding symmetries, (\ref{eq:def}) and (\ref{eq:C}), involve an element $\smash{U_0\in {\cal G}_0}$ which we still need to characterize. This can be done with the help of Weyl transformations whose definition we recall in the next section.\\

\subsection{Defining weights}
Before doing so, it is useful to generalize Eq.~(\ref{eq:canonical}) beyond the adjoint representation. In particular, when diagonalizing the defining action\footnote{The term `defining' is here chosen in place of `fundamental' since there is in general more than one {\it fundamental representation} associated to a given group.} all the $t^j$'s:
\beq
t^j|\rho\rangle=\rho^j|\rho\rangle\,.\label{eq:defining}
\eeq
one obtains the {\it defining weights} $\rho$, which, just as the roots, are real-valued, $(N-1)$-dimensional vectors. For instance, in the SU($2$) case, there are two defining weights $\smash{\rho_1=1/2}$ and $\smash{\rho_2=-1/2}$, while in the SU($3$) case, there are three weights $\smash{\rho_1=(1/2,1/(2\sqrt{3}))}$, $\smash{\rho_2=(-1/2,1/(2\sqrt{3}))}$ and $\smash{\rho_3=(0,-1/\sqrt{3})}$.

The defining weights are closely connected to the roots since the latter arise as all possible differences of two distinct weights $\smash{\alpha_{kl}=\rho_k-\rho_l}$, thus explaining the notation that we introduced above. In the SU(N) case at least, the pair of weights that decompose a given root is unique. Sometimes, given a root $\alpha$, we might want to access the corresponding weights which we denote\footnote{The subscript $\alpha$ should not be confused with the subscript $k$ above. They are related as $\rho_{\alpha_{kl}}=\rho_k$ and $\bar\rho_{\alpha_{kl}}=\rho_l$.} $\rho_\alpha$ and $\bar\rho_\alpha$ such that $\smash{\alpha=\rho_\alpha-\bar\rho_\alpha}$.

Let us finally mention that the SU(N) weights are all such that
\beq
\rho^2=\frac{1}{2}\left(1-\frac{1}{N}\right),
\eeq
whereas
\beq
\rho\cdot\rho'=-\frac{1}{2N}\,,
\eeq
for two distinct weights $\rho$ and $\rho'$, see App.~\ref{app:n}. If follows in particular that the scalar product $\rho\cdot\alpha$ between a defining weight and a root can only take a certain number of values:
\beq
\rho\cdot\alpha & = & +1/2 \quad \mbox{if } \rho=\rho_\alpha\,,\nonumber\\
& = & -1/2 \quad \mbox{if } \rho=\bar\rho_\alpha\,,\nonumber\\
& = & 0 \quad \mbox{otherwise}\,.\label{eq:31}
\eeq 
Similarly, given two roots $\alpha$ and $\beta$, one has
\beq
\alpha\cdot\beta & = & +1 \quad \mbox{if } \beta=\alpha\,,\nonumber\\
& = & -1 \quad \mbox{if } \beta=-\alpha\,,\nonumber\\
& = & -1/2 \quad \mbox{if } \alpha+\beta \mbox{ is a root}\,,\nonumber\\
& = & +1/2 \quad \mbox{if } \alpha-\beta \mbox{ is a root}\,.\label{eq:32}
\eeq

\onecolumngrid
\begin{center}
\begin{figure}[t]
\includegraphics[height=0.14\textheight]{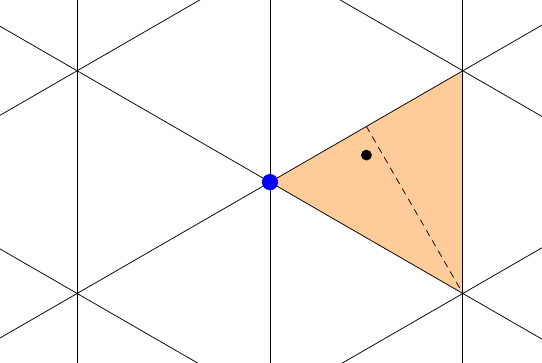}\qquad\includegraphics[height=0.14\textheight]{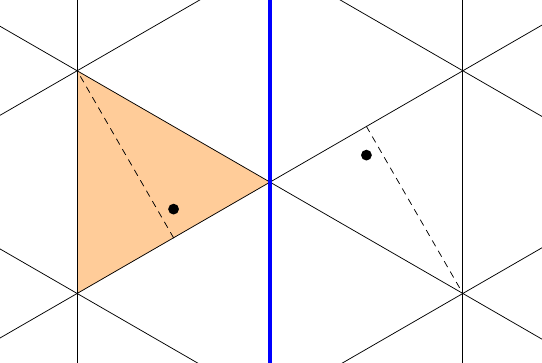}\qquad\includegraphics[height=0.14\textheight]{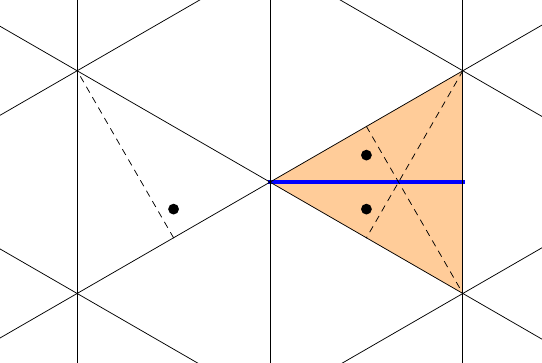}
\caption{Transformation of a Weyl chamber under charge conjugation. The colored chamber represents the various locations of the Weyl chamber along the transformation process. We have chosen a point and a particular axis of the Weyl chamber to ease orientation as the Weyl chamber is transformed. In the first two figures, the blue items represent the transformations that will be applied to the Weyl chamber, $\bar A\to -\bar A^{\rm t}$ and $W_{\alpha_{12}}$ respectively, while in the third figure, the blue item represents the combined effect of these two transformations, which corresponds to a transformation of the original Weyl chamber into itself, more specifically a reflection with respect to its horizontal symmetry axis.}
\label{fig:SU3_C}
\end{figure}
\end{center}
\twocolumngrid

\section{Weyl transformations}\label{sec:weyl}
A {\it Weyl transformation} is a particular element of ${\cal G}_0$. It is a global color transformation associated to a given root $\alpha$ as
\beq
W_\alpha\equiv e^{i\frac{\pi}{\sqrt{2}}(t^\alpha+t^{-\alpha})}e^{i \pi (\rho^j_{\alpha}+\bar\rho^j_{\alpha})t^j}\,.\label{eq:112}
\eeq
Details on the choice of the two factors that enter this definition are given in App.~\ref{app:b} together with a number of properties. In particular we will need to know the adjoint action of $W_\alpha$ on the color algebra.

\subsection{Action on the algebra}
It is shown in App.~\ref{app:b} that
\beq
W_\alpha\,t^j\,W^\dagger_\alpha & = & t^j-2\alpha^j(\alpha\cdot t)\,,\label{eq:241}\\
W_\alpha\,t^\beta\,W^\dagger_\alpha & = & t^{\beta-2(\beta\cdot\alpha)\alpha}\,.\label{eq:242}
\eeq
\noindent{In order to alleviate the notation, we have explicitely used the fact that the SU(N) roots are unit vectors. Otherwise, $\alpha$ in the RHS of Eqs.~(\ref{eq:241}) and (\ref{eq:242}) needs to be replaced by $\alpha/\sqrt{\alpha^2}$. 

We also note that Eq.~(\ref{eq:241}) would remain unchanged were we not to include the second factor in Eq.~(\ref{eq:112}). However, this factor is crucial in order to avoid uninteresting but annoying extra factors in Eq.~(\ref{eq:242}) as we could show in the SU(N) case, see App.~\ref{app:b}. Let us finally mention that the combination $\beta-2(\beta\cdot\alpha)\alpha$ appearing in the RHS of this equation takes different forms depending on the relation between $\alpha$ and $\beta$:}

\beq
\beta-2(\beta\cdot\alpha)\alpha & = & -\alpha \quad \mbox{if } \beta=\alpha\,,\nonumber\\
& = & +\alpha \quad \mbox{if } \beta=-\alpha\,,\nonumber\\
& = & \beta+\alpha \quad \mbox{if } \beta+\alpha \mbox{ is a root}\,,\nonumber\\
& = & \beta-\alpha \quad \mbox{if } \beta-\alpha \mbox{ is a root}\,,\nonumber\\
& = & 0 \quad \mbox{otherwise}\,.\label{eq:321}
\eeq
This combination corresponds to the reflection of $\beta$ with respect to an hyperplane orthogonal to $\alpha$.

\subsection{Weyl chambers}\label{sec:chambers}
The Weyl transformations play an important role in identifying center-symmetric backgrounds among the backgrounds of the form (\ref{eq:bg}). As explained in Ref.~\cite{Reinosa:2020mnx}, one first restricts to the subgroup $\tilde{\cal G}_0$ of transformations of ${\cal G}_0$ that keep the background of the form (\ref{eq:bg}). These transformations are seen to be of the form \cite{Reinosa:2020mnx}
\beq
W V_s(\tau)\,,\label{eq:W}
\eeq
where $W$ is a global color rotation that leaves the diagonal part of the algebra globally invariant, and
\beq
V_s(\tau)\equiv e^{i4\pi\frac{\tau}{\beta} s^jt^j}\,,\label{eq:Vs}
\eeq
with $s$ such that 
\beq
e^{i4\pi s^jt^j}=\mathds{1}\,.
\eeq 
Interestingly, the Weyl transformations $W_\alpha$ provide examples of global color rotation that leave the diagonal part of the algebra globally invariant, as follows from Eq.~(\ref{eq:241}). Similarly, given a root $\alpha$, one always has 
\beq
\smash{e^{i4\pi \alpha^jt^j}=\mathds{1}}\,,
\eeq 
see App.~\ref{app:b}. 

\onecolumngrid
\begin{center}
\begin{figure}[t]
\includegraphics[height=0.16\textheight]{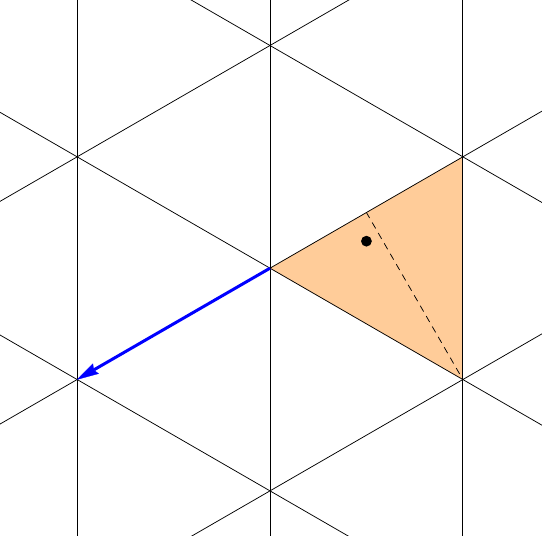}\qquad\includegraphics[height=0.16\textheight]{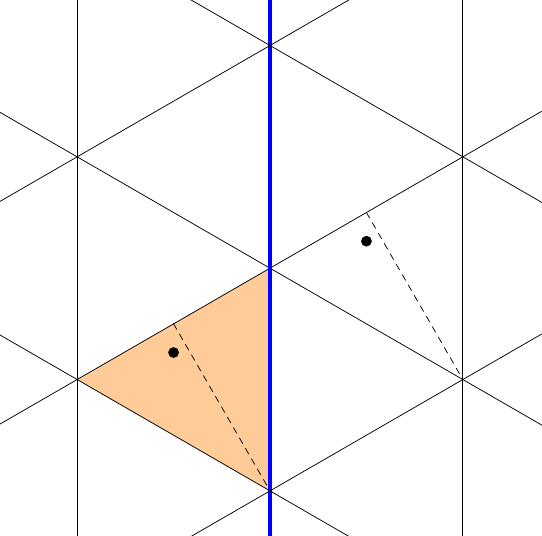}\qquad\includegraphics[height=0.16\textheight]{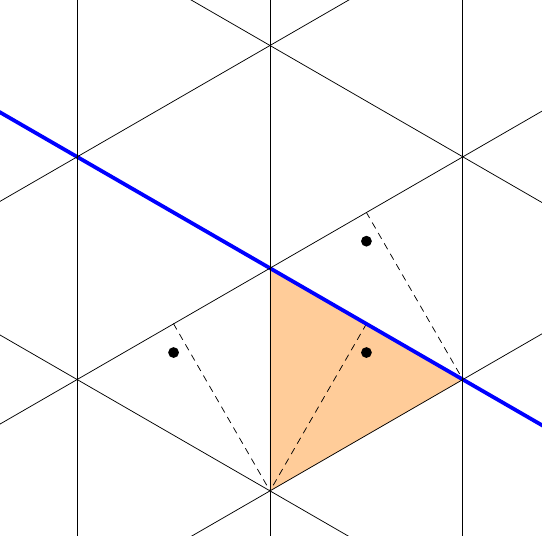}\qquad\includegraphics[height=0.16\textheight]{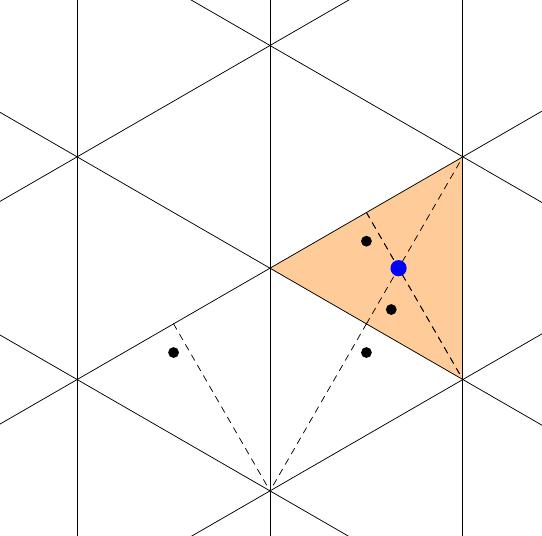}
\caption{Transformation of a Weyl chamber under a center transformation. The colored chamber represents the various locations of the Weyl chamber along the transformation process. We have chosen a point and a particular axis of the Weyl chamber to ease orientation as the Weyl chamber is transformed. In the first three figures, the blue items represent the transformations that will be applied to the Weyl chamber, $V_{-\rho_1}(\tau)$, $W_{\alpha_{12}}$ and $W_{\alpha_{31}}$ respectively, while in the fourth figure, the blue item represents the combined effect of these three transformations, which corresponds to a transformation of the original Weyl chamber into itself, more specifically a rotation by an angle $2\pi/3$ around its center.}
\label{fig:SU3_center}
\end{figure}
\end{center}
\twocolumngrid

Therefore, one can generate all transformations of $\tilde{\cal G}_0$ from the elementary transformations $W_\alpha$ and $\smash{V_\alpha(\tau)}$. In the space of backgrounds of the form (\ref{eq:bg}), the latter correspond to simple geometrical transformations: reflections with respect to hyperplanes orthogonal  to $\alpha$ and translations of $\bar r$ by $4\pi\alpha$. By combining those transformations, one obtains a more interesting generating set, namely the reflections with respect to hyperplanes orthogonal to $\alpha$, displaced by any multiple of $2\pi\alpha$. The benefit of this generating set is that it subdivides the space of backgrounds of the form (\ref{eq:bg}) into regions, known as Weyl chambers, that are connected to each other by elements of $\tilde {\cal G}_0$ and are thus physically equivalent.

In the SU($2$) case, from the roots given above, one obtains that the Weyl chambers are the intervals $\bar r\in [2\pi k, 2\pi (k+1)]$. In the case of SU(3), the Weyl chambers are equilateral triangles, see Fig.~\ref{fig:weyl}. Once the Weyl chambers have been identified, one can easily construct the particular transformations $U_0$ that appear in Eqs.~(\ref{eq:def}) and (\ref{eq:C}). The idea is the same in both cases. A physical transformation such as charge conjugation or an element of $\tilde {\cal G}$ typically displaces a given Weyl chamber. By using elements of $\tilde {\cal G}_0$, one can bring the Weyl chamber back to its original location. In doing so, one generates a transformation of a given Weyl chamber into itself. The fixed points of this transformation correspond to backgrounds obeying (\ref{eq:def}) or (\ref{eq:C}) and the so-constructed combinations of elements of ${\cal G}_0$ provide the transformations $U_0$ appearing in these equations.

\subsection{Charge conjugation}
In the SU($2$) case, the transformation $\bar A_\mu\to -\bar A_\mu^{\rm t}$ is itself an element $i\sigma^2\in{\cal G}_0$. Therefore, Eq.~(\ref{eq:C}) is fulfilled for any choice of background since one can choose $U_0$ to be the inverse of this element. It follows that charge conjugation imposes no constraint in this case. 

In the SU($3$) case, in contrast,  the transformation $\bar A_\mu\to -\bar A_\mu^{\rm t}$ is not an element of ${\cal G}_0$. For backgrounds of the form (\ref{eq:bg}), it corresponds to the transformation $\smash{\bar r\to -\bar r}$. Under this transformation, the Weyl chamber that is highlighted in the first plot of Fig.~\ref{fig:SU3_C} is transformed as shown in the second plot of that same figure. To bring the Weyl chamber back to its original location, one can use the reflection $\smash{(\bar r_3,\bar r_8)\to (-\bar r_3,\bar r_8)}$ which corresponds to $W_{\alpha_{12}}$. From this, not only do we deduce that all backgrounds of the form (\ref{eq:bg}) with $\smash{\bar r=(\bar r^3,0)}$ comply with Eq.~(\ref{eq:C}),\footnote{Since all Weyl chambers are physically equivalent, each Weyl chamber contains one axis of charge-conjugation invariant states.} but we also identify $U_0$ in this equation with $W_{\alpha_{12}}$.

\subsection{Center transformations}
We can proceed similarly to construct the transformation $U_0$ that appears in Eq.~(\ref{eq:def}) or, more directly, the transformation $U_c$ that appears in Eq.~(\ref{eq:def2}). First of all, it can be shown that, modulo elements of $\tilde{\cal G}_0$, the transformations of ${\cal U}_1$ that leave the Cartan subalgebra globally invariant are {\it winding transformations} of the form $V_{-\rho}(\tau)$, with $\rho$ one of the defining weights of the algebra.\footnote{Note that, unlike $V_{\alpha} (\tau)$, $V_{-\rho}(\tau)$ does not obey the condition underneath eq. \eqref{eq:Vs} and is therefore not a part of $\mathcal{G}_0$.} The winding transformations act on the algebra as
\beq
V_{-\rho}(\tau)\,t^j\,V^\dagger_{-\rho}(\tau) & = & t^j\,,\label{eq:18}\\
V_{-\rho}(\tau)\,t^\alpha\,V^\dagger_{-\rho}(\tau) & = & e^{-i4\pi\frac{\tau}{\beta}\rho\cdot\alpha}t^\alpha\,,\label{eq:19}
\eeq
where we have used similar considerations as in Eq.~(\ref{eq:Ut}). In the space of backgrounds of the form (\ref{eq:bg}), they correspond to translations by $-4\pi\rho$.  

In the SU($2$) case for instance, if we choose to work on the Weyl chamber $[0,2\pi]$, we see that $V_{-\rho_1}(\tau)$ transforms it into $[-2\pi,0]$. This Weyl chamber can be brought back to its original location by applying $W_{\alpha_{12}}$. Eventually, this produces the transformation $\bar r\to 2\pi-\bar r$ that leaves the original Weyl chamber globally invariant and identifies $\bar r=\pi$ as a fixed point. It follows that the transformation $U_c$ that appears in (\ref{eq:def2}) is the transformation
\beq
U_c=W_{\alpha_{12}}V_{-\rho_1}(\tau)\,,\label{eq:t2}
\eeq
and that the center-symmetric background is indeed (\ref{eq:pi}). As for the SU($3$) case, a similar argumentation leads to
\beq
U_c=W_{\alpha_{31}}W_{\alpha_{12}}V_{-\rho_1}(\tau)\,,\label{eq:t3}
\eeq
corresponding to the center-symmetric background (\ref{eq:4pio3}). For a graphical representation of this construction, see Fig.~\ref{fig:SU3_center}. We mention that simple rules to permute the order between the various $W_\alpha$ or between the $W_\alpha$ and $V_\rho$ are provided in App.~\ref{app:b}.
 
We are now fully equipped to investigate the constraints that charge conjugation and center symmetry impose on correlation functions.

\section{Charge conjugation constraints}\label{sec:C_constraints}
As already mentioned above, charge conjugation imposes no constraints in the SU($2$) case. 

In the SU($3$) case, we have seen that, for backgrounds of the form (\ref{eq:bg}) with $\smash{\bar r=(\bar r^3,0)}$, the transformation $\bar A_\mu\to -\bar A_\mu^{\rm t}$ combined with $W_{\alpha_{12}}$ is a symmetry. Using Eqs.~(\ref{eq:241})-(\ref{eq:242}), we find that this symmetry acts on the generators of the algebra as
\beq
t^{0^{(3)}}\to t^{0^{(3)}}\,, \quad t^{0^{(8)}}\to -t^{0^{(8)}}\,,
\eeq
and
\beq
t^{\pm\alpha_{12}} & \to & -t^{\pm\alpha_{12}}\,,\nonumber\\
t^{\pm\alpha_{23}} & \to & -t^{\pm\alpha_{31}}\,,\\
t^{\pm\alpha_{31}} & \to & -t^{\pm\alpha_{23}}\,.\nonumber
\eeq
We can further use color rotation invariance and redefine the above transformation such that it appears as\footnote{To this purpose, we consider the color rotation $e^{i2\pi\rho_3^jt^j}$ and exploit the fact that $\smash{\rho_3\cdot\alpha_{12}=0}$, $\smash{\rho_3\cdot\alpha_{23}=-1/2}$ and $\smash{\rho_3\cdot\alpha_{31}=+1/2}$.}
\beq
t^{\pm\alpha_{12}} & \to & -t^{\pm\alpha_{12}}\,,\nonumber\\
t^{\pm\alpha_{23}} & \to & t^{\pm\alpha_{31}}\,,\\
t^{\pm\alpha_{31}} & \to & t^{\pm\alpha_{23}}\,.\nonumber
\eeq
It follows that the correlation/vertex functions are invariant under the transformation
\beq
A_\mu^3\to A_\mu^3\,, \quad A_\mu^8\to -A_\mu^8\,,
\eeq
and
\beq
A_\mu^{\pm\alpha_{12}}  & \to & -A_\mu^{\pm\alpha_{12}}\,,\nonumber\\
A_\mu^{\pm\alpha_{23}}  & \to & A_\mu^{\pm\alpha_{31}}\,,\label{eq:charge3}\\
A_\mu^{\pm\alpha_{31}}  & \to & A_\mu^{\pm\alpha_{23}}\,,\nonumber
\eeq
corresponding to a change of sign of the components in the color $0^{(8)}$ and $\pm\alpha_{12}$ directions and a permutation of the components in the $\pm\alpha_{23}$ and $\pm\alpha_{31}$ directions.

Let us now investigate the consequences of this symmetry on the correlation functions, first using some examples and then in full generality.

\subsection{Some examples}

\subsubsection{Neutral sector}
Consider a correlation function whose external legs are all in the neutral sector:
\beq
\langle A^3_{\mu_1}\cdots A_{\mu_m}^3A_{\nu_1}^8\cdots A_{\nu_n}^8\rangle\,.
\eeq
The constraints from charge conjugation invariance read
\beq
& & \langle A^3_{\mu_1}\cdots A_{\mu_m}^3A_{\nu_1}^8\cdots A_{\nu_n}^8\rangle\nonumber\\
& & \hspace{1.0cm}=\,(-1)^n\langle A^3_{\mu_1}\cdots A_{\mu_m}^3A_{\nu_1}^8\cdots A_{\nu_n}^8\rangle\,.
\eeq
As a consequence, all correlators with an odd number of components along the color $8$ direction need to vanish:
\beq
\langle A^3_{\mu_1}\cdots A_{\mu_m}^3A_{\nu_1}^8\cdots A_{\nu_{2p+1}}^8\rangle=0\,.
\eeq
 We expect these constraints to always be valid since charge conjugation should not be spontaneously broken.

\subsubsection{Two-point function}
In particular, for the two-point function, it follows that
\beq
{\cal G}_{\mu\nu}^{0^{(3)}0^{(8)}}={\cal G}_{\mu\nu}^{0^{(8)}0^{(3)}}=0\,.
\eeq
Combined with Eq.~(\ref{eq:c1}) and (\ref{eq:c2}), this implies
\beq
{\cal G}_{\mu\nu}^{\kappa\lambda}={\cal G}_{\mu\nu}^\lambda\delta^{\kappa(-\lambda)}\,,
\eeq
as in the SU($2$) case. We mention, however, that we do not know at this point whether and how the neutral diagonal elements ${\cal G}_{\mu\nu}^{0^{(3)}0^{(3)}}$ and ${\cal G}_{\mu\nu}^{0^{(8)}0^{(8)}}$ are connected to each other. We will come back to this question below as it is linked to center symmetry.

In the charged sector, we have that
\beq
{\cal G}_{\mu\nu}^{\pm\alpha_{23}}={\cal G}_{\mu\nu}^{\pm\alpha_{31}}\,,
\eeq
as follows from (\ref{eq:charge3}) but we cannot tell at this point whether and how they are connected to the components along the $\pm\alpha_{12}$ directions. Again, we will come back to this question below.

\subsubsection{Three-point function}
For the three-point functions, we have
\beq
\langle A_\mu^3 A_\nu^3 A_\rho^8\rangle=\langle A_\mu^8 A_\nu^8 A_\rho^8\rangle=0\,.
\eeq
Similarly
\beq
\langle A_\mu^8 A_\nu^{\alpha_{12}}A_\rho^{-\alpha_{12}}\rangle=0\,,\label{eq:500}
\eeq
as well as
\beq
\langle A_\mu^3 A_\nu^{\alpha_{23}}A_\rho^{-\alpha_{23}}\rangle=\langle A_\mu^3 A_\nu^{\alpha_{31}}A_\rho^{-\alpha_{31}}\rangle\,,
\eeq
and
\beq
\langle A_\mu^8 A_\nu^{\alpha_{23}}A_\rho^{-\alpha_{23}}\rangle=-\langle A_\mu^8 A_\nu^{\alpha_{31}}A_\rho^{-\alpha_{31}}\rangle\,.\label{eq:502}
\eeq
Finally
\beq
\langle A_\mu^{\alpha_{12}} A_\nu^{\alpha_{23}}A_\rho^{\alpha_{31}}\rangle=-\langle A_\mu^{\alpha_{12}} A_\nu^{\alpha_{31}}A_\rho^{\alpha_{23}}\rangle\,.\label{eq:503}
\eeq

\subsection{General case}
To derive the general constraints from charge conjugation invariance, it is convenient to introduce the field $\smash{A_\mu^\sigma\equiv A_\mu^3+\sigma i A_\mu^8}$ (with $\smash{\sigma=\pm 1}$) which transforms as
\beq
A_\mu^\sigma \to A_\mu^{-\sigma}\,.
\eeq
A general correlation function takes then the form
\beq
& & \Big\langle A_{\mu_1}^+\cdots A_{\mu_m}^+A_{\nu_1}^-\cdots A_{\nu_n}^-\nonumber\\
& & \hspace{0.2cm}\,A_{\rho_1}^{\alpha_{12}}\cdots A_{\rho_p}^{\alpha_{12}}A_{\sigma_1}^{-\alpha_{12}}\cdots A_{\sigma_q}^{-\alpha_{12}}\nonumber\\
& & \hspace{0.2cm}\,A_{\omega_1}^{\alpha_{23}}\cdots A_{\omega_k}^{\alpha_{23}}A_{\tau_1}^{-\alpha_{23}}\cdots A_{\tau_\ell}^{-\alpha_{23}}\nonumber\\
& & \hspace{0.2cm}A_{\eta_1}^{\alpha_{31}}\cdots A_{\eta_i}^{\alpha_{31}}A_{\zeta_1}^{-\alpha_{31}}\cdots A_{\zeta_j}^{-\alpha_{31}}\Big\rangle\,,\label{eq:general}
\eeq
which we write formally as
\beq
\langle (+)_{mn}(12)_{pq}(23)_{k\ell}(31)_{ij}\rangle\,.
\eeq
We note that we have the constraint
\beq
(p-q)\alpha_{12}+(k-\ell)\alpha_{23}+(i-j)\alpha_{31}=0\,,
\eeq
as follows from color conservation, see the discussion below Eq.~(\ref{eq:color_constraint}). Now, because $\alpha_{12}+\alpha_{23}+\alpha_{31}=0$, this rewrites as $\smash{(p+j-q-i)\alpha_{12}+(k+j-\ell-i)\alpha_{23}=0}$. Since $\alpha_{12}$ and $\alpha_{23}$ are linearly independent, this eventually leads to $\smash{p-q=i-j=k-\ell}$. Moreover, correlation functions involving $A_\mu^3$ and $A_\mu^8$ rather than $A_\mu^\pm$ can be obtained through appropriate linear combinations of (\ref{eq:general}).

With this compact notation, the constraints due to charge conjugation invariance read
\beq
& & \langle (+)_{mn}(12)_{pq}(23)_{k\ell}(31)_{ij}\rangle\nonumber\\
& & \hspace{0.5cm}=\,(-1)^{p+q}\langle (-)_{mn}(12)_{pq}(31)_{k\ell}(23)_{ij}\rangle\,.
\eeq
One easily checks that this identity contains the constraints already derived above and allows one to generate all other possible constraints related to charge conjugation invariance.

\section{$Z_2$-symmetry constraints}\label{sec:Z2_constraints}

Using Eqs.~(\ref{eq:241})-(\ref{eq:321}) and Eqs.~(\ref{eq:18})-(\ref{eq:19}), we find that the action of $U_c$ on the SU($2$) algebra reads
\beq
U_c\,t^j\,U^\dagger_c=-t^j\,,
\eeq
as well as
\beq
U_c\,t^{\pm\alpha_{12}}\,U^\dagger_c & = & e^{\mp i2\pi\frac{\tau}{\beta}}\,t^{\mp\alpha_{12}}\,,
\eeq
where we have used that $\rho_1\cdot\alpha_{12}=1/2$. From this, one reads the corresponding ${\cal U}_c$ and deduces that the correlation functions are invariant under the transformation
\beq
\delta A_\mu^3 & \to & -\delta A_\mu^3\,,\\
\delta A_\mu^{\pm\alpha_{12}} & \to & e^{\pm i2\pi \frac{\tau_\mu}{\beta}}\delta A_\mu^{\mp\alpha_{12}}\,,\label{eq:Z2charged}
\eeq
where $\delta A$ was defined below Eq.~(\ref{eq:cntd}) and $\tau_\mu$ stands for the Euclidean time argument associated to the index $\mu$. Let us now analyze the consequences of this symmetry on the correlation functions, first using some examples, and then in full generality.

\subsection{Some examples}

\subsubsection{Neutral sector}
If we consider correlation functions that are purely in the neutral sector, the constraint (\ref{eq:constraints}) takes the form
\beq
\langle A_{\mu_1}^3\cdots A_{\mu_n}^3\rangle=(-1)^n\langle A_{\mu_1}^3\cdots A_{\mu_n}^3\rangle\,.
\eeq
Thus, functions with an even number of external legs are unconstrained, while those with an odd number of external legs should vanish
\beq
\langle A_{\mu_1}^3\cdots A_{\mu_{2p+1}}^3\rangle=0\,,
\eeq
as long as center symmetry is not broken. 

\subsubsection{Two-point function}
In particular there is no constraint on the two-point function in the neutral sector.\footnote{In Refs.~\cite{vanEgmond:2021jyx,vanEgmond:2022nuo}, we have seen that the two-point function in the neutral sector develops a zero-mode at the deconfinement transition. Although this is a combined consequence of center symmetry and of the second order nature of the transition in the SU($2$) case, it is not of the same type than the symmetry constraints that we are presently discussing. The latter apply indeed over the whole confining phase and not just at the transition.} 

On the contrary, from Eq.~(\ref{eq:Z2charged}), we find the following constraint on the two-point function in the charged sector
\beq
& & \langle A_{\mu}^{\alpha_{12}} A_{\nu}^{-\alpha_{12}}\rangle=e^{i2\pi\frac{\tau_\mu-\tau_\nu}{\beta}}\langle A_{\mu}^{-\alpha_{12}} A_{\nu}^{\alpha_{12}}\rangle\,,
\eeq
where $\tau_\mu$ and $\tau_\nu$ are the Euclidean time arguments associated with the fields carrying the indices $\mu$ and $\nu$ respectively. In other words (we now make the position arguments explicit)
\beq
{\cal G}^{\alpha_{12}(-\alpha_{12})}_{\mu\nu}(x-y)=e^{i2\pi\frac{\tau_x-\tau_y}{\beta}}{\cal G}^{(-\alpha_{12})\alpha_{12}}_{\mu\nu}(x-y)\,.\nonumber\\
\eeq
In Fourier space, this means that ${\cal G}_{\mu\nu}^{\alpha_{12}(-\alpha_{12})}(Q)$ and ${\cal G}_{\mu\nu}^{(-\alpha_{12})\alpha_{12}}(Q)$ are related by a frequency shift of $2\pi T$. With our convention $\partial_\mu\to -iQ_\mu$ for the Fourier transform, this reads
\beq
{\cal G}^{\alpha_{12}(-\alpha_{12})}_{\mu\nu}(Q)={\cal G}^{(-\alpha_{12})\alpha_{12}}_{\mu\nu}(Q+2\pi T N)\,,\label{eq:z2}
\eeq
with $\smash{N=(1,\vec{0})}$. Any violation of this identity signals a breaking of center symmetry.

\subsubsection{Three-point function}
For the three-point function, either all color directions are neutral and then
\beq
\langle A_\mu^3(x)A_\nu^3(y)A_\rho^3(z)\rangle=0\,,\label{eq:z21}
\eeq
or, only one is neutral and then
\beq
& & \langle A_\mu^{\alpha_{12}}(x)A_\nu^{-\alpha_{12}}(y)A_\rho^3(z)\rangle\nonumber\\
& & \hspace{0.5cm}=\,-e^{i2\pi\frac{\tau_\mu-\tau_\nu}{\beta}}\langle A_\mu^{-\alpha_{12}}(x)A_\nu^{\alpha_{12}}(y)A_\rho^3(z)\rangle\,.
\eeq
 In Fourier space, this leads to
\beq
& & \langle A_\mu^{\alpha_{12}}(P)A_\nu^{-\alpha_{12}}(Q)A_\rho^3(K)\rangle\nonumber\\
& & \hspace{0.5cm}=\,-\langle A_\mu^{-\alpha_{12}}(P+2\pi TN)A_\nu^{\alpha_{12}}(Q-2\pi TN)A_\rho^3(K)\rangle\,.\label{eq:z22}\nonumber\\
\eeq

\subsection{General case}
A general SU($2$) correlation function takes the form
\beq
& & \langle A_{\mu_1}^3\cdots A_{\mu_n}^3A_{\rho_1}^{\alpha_{12}}\cdots A_{\rho_p}^{\alpha_{12}}A_{\sigma_1}^{-\alpha_{12}}\cdots A_{\sigma_p}^{-\alpha_{12}}\rangle
\eeq
which we denote more simply as $\langle (3)_n(12)_p\rangle$. The constraints due to center symmetry read
\beq
\langle (3)_n(12)_p\rangle & = & (-1)^ne^{i2\pi \frac{\tau_{\rho_1}+\cdots+\tau_{\rho_p}-\tau_{\sigma_1}-\cdots\tau_{\sigma_p}}{\beta}}\langle (3)_n(21)_p\rangle\,.\label{eq:z2n}\nonumber\\
\eeq
We of course retrieve the previously obtained constraints as particular cases of this general identity.

\section{$Z_3$-symmetry constraints}\label{sec:Z3_constraints}

Using Eqs.~(\ref{eq:241})-(\ref{eq:321}) and Eqs.~(\ref{eq:18})-(\ref{eq:19}), we find again that the neutral and charged sectors decouple under the adjoint action of $U_c$. In the neutral sector, the transformation corresponds to a rotation by an angle $2\pi/3$, whereas in the charged sector, we find
\beq
U_c\,t^{\pm\alpha_{12}}\,U_c^\dagger & = & e^{\mp i2\pi\frac{\tau}{\beta}}t^{\pm\alpha_{23}}\,,\nonumber\\
U_c\,t^{\pm\alpha_{23}}\,U_c^\dagger & = & t^{\pm\alpha_{31}}\,,\\
U_c\,t^{\pm\alpha_{31}}\,U_c^\dagger & = & e^{\pm i2\pi\frac{\tau}{\beta}}t^{\pm\alpha_{12}}\,.\nonumber
\eeq
In terms of the gauge field, the rotation by an angle $2\pi/3$ in the neutral sector reads
\beq
\delta A_\mu^\sigma \to e^{i\frac{2\pi}{3}\sigma}\delta A_\mu^\sigma\,,
\eeq
while in the charged sector, we have
\beq
\delta A_\mu^{\pm\alpha_{12}} & \to & e^{\pm i2\pi \frac{\tau_\mu}{\beta}}\delta A_\mu^{\pm\alpha_{31}}\,,\nonumber\\
\delta A_\mu^{\pm\alpha_{23}} & \to & e^{\mp i2\pi\frac{\tau_\mu}{\beta}}\delta A_\mu^{\pm\alpha_{12}}\,,\\
\delta A_\mu^{\pm\alpha_{31}} & \to & \delta A_\mu^{\pm\alpha_{23}}\,,\nonumber
\eeq
that is a permutation $12\to 31\to 23\to 12$ with appropriate phase factors. Let us now analyze the consequences of this symmetry on the correlation functions.

\subsection{Some examples}

\subsubsection{Neutral sector}
If we consider correlation functions that are purely in the neutral sector, the constraint (\ref{eq:constraints}) takes the form
\beq
\langle A_{\mu_1}^{\sigma^1}\cdots A_{\mu_n}^{\sigma^n}\rangle=e^{i\frac{2\pi}{3}(\sigma^1+\cdots+\sigma^n)}\langle A_{\mu_1}^{\sigma^1}\cdots A_{\mu_n}^{\sigma^n}\rangle\,.\nonumber\\
\eeq
Thus, correlation functions such that $\sigma^1+\cdots+\sigma^n\notin 3\mathds{Z}$ need to vanish if the center symmetry is not broken.

\subsubsection{Two-point function}
 For the two-point function, without loss of generality,\footnote{Indeed, $\langle A^-_\mu A^-_\nu\rangle$ is nothing but the complex conjugate of $\langle A^+_\mu A^+_\nu\rangle$.} we can consider $\langle A^+_\mu A^+_\nu\rangle$ and then
\beq
0=\langle A^+_\mu A^+_\nu\rangle & = & \langle A^3_\mu A^3_\nu\rangle-\langle A^8_\mu A^8_\nu\rangle\nonumber\\
& +i & \Big[\langle A^3_\mu A^8_\nu \rangle+\langle A^8_\mu A^3_\nu\rangle\Big]\,.
\eeq
Taking the real and imaginary parts of this identity, we find that both
\beq
\langle A_\mu^3A_\nu^3\rangle-\langle A_\mu^8A_\nu^8\rangle=0\,\label{eq:z3}
\eeq
and 
\beq
\langle A_\mu^3A_\nu^8\rangle+\langle A_\mu^8A_\nu^3\rangle=0\,,\label{eq:first}
\eeq
if center symmetry is not broken.

In fact, the second identity is always fulfilled due to the constraints from charge conjugation invariance, see Sec.~\ref{sec:C_constraints}. On the other hand, the first combination has no reason to remain $0$ if center symmetry is broken. One could invoke color rotation invariance but the fact that the gauge fixing introduces a preferred color direction along $\lambda_3/2$ prevents us from doing so. We deduce that this second combination can be used as an (infinite collection of) order parameter(s) for center-symmetry. We have tested this hypothesis in Ref.~\cite{vanEgmond:2022nuo} for the case of the chromo-electric component of the propagator in the zero-frequency limit. We now see that this should apply to the chromo-magnetic component as well and for any value of the external momentum. This will be studied in a future work.

In the charged sector, we find
\beq
\langle A_{\mu}^{\alpha_{12}} A_{\nu}^{-\alpha_{12}}\rangle  & = & e^{i2\pi\frac{\tau_\mu-\tau_\nu}{\beta}}\langle A_{\mu}^{\alpha_{31}} A_{\nu}^{-\alpha_{31}}\rangle\nonumber\\
& = & e^{i2\pi\frac{\tau_\mu-\tau_\nu}{\beta}}\langle A_{\mu}^{\alpha_{23}} A_{\nu}^{-\alpha_{23}}\rangle\,,
\eeq
that is
\beq
& & {\cal G}^{\alpha_{12}(-\alpha_{12})}_{\mu\nu}(x-y)\nonumber\\
& & \hspace{1.0cm}=\,e^{i2\pi\frac{\tau_x-\tau_y}{\beta}}{\cal G}^{\alpha_{23}(-\alpha_{23})}_{\mu\nu}(x-y)\nonumber\\
& & \hspace{1.0cm}=\,e^{i2\pi\frac{\tau_x-\tau_y}{\beta}}{\cal G}^{\alpha_{31}(-\alpha_{31})}_{\mu\nu}(x-y)\,.
\eeq
In Fourier space, this means that ${\cal G}_{\mu\nu}^{\alpha_{12}(-\alpha_{12})}(Q)$ is related to ${\cal G}_{\mu\nu}^{\alpha_{23}(-\alpha_{23})}(Q)$ and ${\cal G}_{\mu\nu}^{\alpha_{31}(-\alpha_{31})}(Q)$ by a mere shift of the external frequency by $2\pi T$. With our convention $\partial_\mu\to -iQ_\mu$ for the Fourier transform, this reads
\beq
{\cal G}^{\alpha_{12}(-\alpha_{12})}_{\mu\nu}(Q) & = & {\cal G}^{\alpha_{23}(-\alpha_{23})}_{\mu\nu}(Q+2\pi T N)\nonumber\\
& = & {\cal G}^{\alpha_{31}(-\alpha_{31})}_{\mu\nu}(Q+2\pi T N)\,,\label{eq:z30}
\eeq
with $\smash{N=(1,\vec{0})}$. Any violation of these identities signals a breaking of center symmetry. On the other hand, the fact that ${\cal G}_{\mu\nu}^{\alpha_{23}(-\alpha_{23})}$ and ${\cal G}_{\mu\nu}^{\alpha_{31}(-\alpha_{31})}$ agree with each other is a consequence of charge conjugation invariance, as we have seen in Sec.~\ref{sec:C_constraints}.

\subsubsection{Three-point function}
Consider first the case where all the external legs are in the neutral sector. Without loss of generality, we can consider $\langle A_\mu^+A_\nu^+A_\rho^-\rangle$ as the other cases are obtained from permutations or complex conjugation. We then find
\beq
0 & = & \langle A_\mu^+A_\nu^+A_\rho^-\rangle\nonumber\\
& = & \langle A_\mu^3A_\nu^3A_\rho^3\rangle-\langle A_\mu^8A_\nu^8A_\rho^3\rangle\nonumber\\
& + & \langle A_\mu^3A_\nu^8A_\rho^8\rangle+\langle A_\mu^8A_\nu^3A_\rho^8\rangle\nonumber\\
& + & i\Big[\langle A_\mu^8A_\nu^8A_\rho^8\rangle-\langle A_\mu^3A_\nu^3A_\rho^8\rangle\nonumber\\
& + & \langle A_\mu^8A_\nu^3A_\rho^3\rangle+\langle A_\mu^3A_\nu^8A_\rho^3\rangle\Big].
\eeq
Taking the real and imaginary parts this gives
\beq
0 & = & \langle A_\mu^3A_\nu^3A_\rho^3\rangle-\langle A_\mu^8A_\nu^8A_\rho^3\rangle\nonumber\\
& + & \langle A_\mu^3A_\nu^8A_\rho^8\rangle+\langle A_\mu^8A_\nu^3A_\rho^8\rangle
\eeq
and
\beq
0 & = & \langle A_\mu^8A_\nu^8A_\rho^8\rangle-\langle A_\mu^3A_\nu^3A_\rho^8\rangle\nonumber\\
& + & \langle A_\mu^8A_\nu^3A_\rho^3\rangle+\langle A_\mu^3A_\nu^8A_\rho^3\rangle\,.
\eeq
Writing similar formulas for permutations of $(x,y,z)$ and $(\mu,\nu,\rho)$, we find that these relations are equivalent to
\beq
\langle A_\mu^3A_\nu^3A_\rho^3\rangle & = & -\langle A_\mu^8A_\nu^8A_\rho^3\rangle\nonumber\\
& = & -\langle A_\mu^8A_\nu^3A_\rho^8\rangle\nonumber\\
& = & -\langle A_\mu^3A_\nu^8A_\rho^8\rangle\,,\label{eq:z31}
\eeq
and
\beq
\langle A_\mu^8A_\nu^8A_\rho^8\rangle & = & -\langle A_\mu^3A_\nu^3A_\rho^8\rangle\nonumber\\
& = & -\langle A_\mu^3A_\nu^8A_\rho^3\rangle\nonumber\\
& = & -\langle A_\mu^8A_\nu^3A_\rho^3\rangle\,.
\eeq
A priori each of these identities could be used as a probe for the deconfinement transition. However, the second set is always (trivially) fulfilled for our particular choice of background due to charge conjugation invariance. 

Consider now a three-point function involving charged modes. Let us first consider the correlators $\langle A_\mu^\alpha A_\nu^{-\alpha}A_\rho^3\rangle$ and $\langle A_\mu^\alpha A_\nu^{-\alpha}A_\rho^8\rangle$ which we combine into $\langle A_\mu^\alpha A_\nu^{-\alpha}A_\rho^\sigma\rangle$. We find
\beq
& & \langle A_\mu^{\alpha_{12}}A_\nu^{-\alpha_{12}} A_\rho^\sigma\rangle\nonumber\\
& & \hspace{1.0cm}=\,e^{i2\pi\frac{\tau_\mu-\tau_\nu}{\beta}+i\frac{2\pi}{3}\sigma}\langle A_\mu^{\alpha_{31}}A_\nu^{-\alpha_{31}} A_\rho^\sigma\rangle\,,\label{eq:93}\\
& & \langle A_\mu^{\alpha_{23}}A_\nu^{-\alpha_{23}} A_\rho^\sigma\rangle\nonumber\\
& & \hspace{1.0cm}=\,e^{-i2\pi\frac{\tau_\mu-\tau_\nu}{\beta}+i\frac{2\pi}{3}\sigma}\langle A_\mu^{\alpha_{12}}A_\nu^{-\alpha_{12}} A_\rho^\sigma\rangle\,,\\
& & \langle A_\mu^{\alpha_{31}}A_\nu^{-\alpha_{31}} A_\rho^\sigma\rangle\nonumber\\
& & \hspace{1.0cm}=\,e^{+i\frac{2\pi}{3}\sigma}\langle A_\mu^{\alpha_{23}}A_\nu^{-\alpha_{23}} A_\rho^\sigma\rangle\,,\label{eq:95}
\eeq
which can interpret as two independent equations giving two correlation functions in terms of the third one. Equivalently, this rewrites
\beq
& & \langle A_\mu^{\alpha_{23}}A_\nu^{-\alpha_{23}} A_\rho^3\rangle\nonumber\\
& & \hspace{1.0cm}=\,-\frac{1}{2}e^{-i2\pi\frac{\tau_\mu-\tau_\nu}{\beta}}\langle A_\mu^{\alpha_{12}}A_\nu^{-\alpha_{12}} A_\rho^3\rangle\nonumber\\
& & \hspace{1.0cm}-\,\frac{\sqrt{3}}{2}e^{-i2\pi\frac{\tau_\mu-\tau_\nu}{\beta}}\langle A_\mu^{\alpha_{12}}A_\nu^{-\alpha_{12}} A_\rho^8\rangle\,,
\eeq
\beq
& & \langle A_\mu^{\alpha_{31}}A_\nu^{-\alpha_{31}} A_\rho^3\rangle\nonumber\\
& & \hspace{1.0cm}=\,-\frac{1}{2}e^{-i2\pi\frac{\tau_\mu-\tau_\nu}{\beta}}\langle A_\mu^{\alpha_{12}}A_\nu^{-\alpha_{12}} A_\rho^3\rangle\nonumber\\
& & \hspace{1.0cm}+\,\frac{\sqrt{3}}{2}e^{-i2\pi\frac{\tau_\mu-\tau_\nu}{\beta}}\langle A_\mu^{\alpha_{12}}A_\nu^{-\alpha_{12}} A_\rho^8\rangle\,,
\eeq
\beq
& & \langle A_\mu^{\alpha_{23}}A_\nu^{-\alpha_{23}} A_\rho^8\rangle\nonumber\\
& & \hspace{1.0cm}=\,-\frac{1}{2}e^{-i2\pi\frac{\tau_\mu-\tau_\nu}{\beta}}\langle A_\mu^{\alpha_{12}}A_\nu^{-\alpha_{12}} A_\rho^8\rangle\nonumber\\
& & \hspace{1.0cm}+\,\frac{\sqrt{3}}{2}e^{-i2\pi\frac{\tau_\mu-\tau_\nu}{\beta}}\langle A_\mu^{\alpha_{12}}A_\nu^{-\alpha_{12}} A_\rho^3\rangle\,,
\eeq
and
\beq
& & \langle A_\mu^{\alpha_{31}}A_\nu^{-\alpha_{31}} A_\rho^8\rangle\nonumber\\
& & \hspace{1.0cm}=\,-\frac{1}{2}e^{-i2\pi\frac{\tau_\mu-\tau_\nu}{\beta}}\langle A_\mu^{\alpha_{12}}A_\nu^{-\alpha_{12}} A_\rho^8\rangle\nonumber\\
& & \hspace{1.0cm}-\,\frac{\sqrt{3}}{2}e^{-i2\pi\frac{\tau_\mu-\tau_\nu}{\beta}}\langle A_\mu^{\alpha_{12}}A_\nu^{-\alpha_{12}} A_\rho^3\rangle.
\eeq
Upon using charge conjugation invariance, the only non-trivial information that arises from center symmetry is
\beq
& & \langle A_\mu^{\alpha_{23}}A_\nu^{-\alpha_{23}} A_\rho^3\rangle\nonumber\\
& & \hspace{1.0cm}=\,-\frac{1}{2}e^{-i2\pi\frac{\tau_\mu-\tau_\nu}{\beta}}\langle A_\mu^{\alpha_{12}}A_\nu^{-\alpha_{12}} A_\rho^3\rangle\label{eq:z32}
\eeq
and
\beq
& & \langle A_\mu^{\alpha_{23}}A_\nu^{-\alpha_{23}} A_\rho^8\rangle\nonumber\\
& & \hspace{1.0cm}=\,\frac{\sqrt{3}}{2}e^{-i2\pi\frac{\tau_\mu-\tau_\nu}{\beta}}\langle A_\mu^{\alpha_{12}}A_\nu^{-\alpha_{12}} A_\rho^3\rangle\,,\label{eq:z33}
\eeq
since the other correlators are fixed through Eqs.~(\ref{eq:500})-(\ref{eq:502}).

Finally we find the constraint
\beq
\langle A_\mu^{\alpha_{12}} A_\nu^{\alpha_{23}}A_\rho^{\alpha_{31}}\rangle=e^{i2\pi\frac{\tau_\mu-\tau_\nu}{\beta}}\langle A_\mu^{\alpha_{31}} A_\nu^{\alpha_{12}}A_\rho^{\alpha_{23}}\rangle\,.\label{eq:z34}
\eeq

\subsection{General case}
The argument in the previous section is in fact more general. 

The symmetry will always connect the correlators $\langle (+)_{mn}(12)_{pq}(23)_{k\ell}(31)_{ij}\rangle$ and $\langle (+)_{mn}(23)_{pq}(31)_{k\ell}(12)_{ij}\rangle$. More precisely, we find
\beq
& & \langle (+)_{mn}(23)_{pq}(31)_{k\ell}(12)_{ij}\rangle\nonumber\\
& & \hspace{1.0cm}=\,e^{i2\pi\frac{\tau_i+\tau_q-\tau_j-\tau_p}{\beta}+i\frac{2\pi}{3}(m-n)\sigma}\nonumber\\
& & \hspace{1.5cm}\times\,\langle (+)_{mn}(12)_{pq}(23)_{k\ell}(31)_{ij}\rangle\,,
\eeq
where $\tau_p$ (resp. $\tau_q$) is the sum of the time arguments associated to the field along the color $\alpha_{12}$ (resp. $(-\alpha_{12})$) direction, $\tau_k$ (resp. $\tau_\ell$) is the sum of the time arguments associated to the field along the color $\alpha_{23}$ (resp. $(-\alpha_{23})$) direction, and $\tau_i$ (resp. $\tau_j$) is the sum of the time arguments associated to the field along the color $\alpha_{31}$ (resp. $(-\alpha_{31})$) direction.

\pagebreak

\section{SU(N) case}
Let us now see how the previous considerations extend to the SU(N) case.

\subsection{Weyl chambers}
Recall that the Weyl chambers appear as the result of the tiling of the Cartan subalgebra by the network of hyperplanes orthogonal to the roots and displaced from the origin by any multiple of $2\pi$ times the corresponding root, see Sec.~\ref{sec:chambers}. To construct the Weyl chambers more explicitly, one can proceed as follows. 

First, one selects $N-1$ out of the $N$ defining weights. We denote them as $\rho_1,\dots,\rho_{N-1}$.\footnote{This labelling does not need to be the particular one chosen in Eq.~(\ref{eq:rho}).} It is easily seen that they form a basis \cite{vanEgmond:2021wlt}. Next, denoting the remaining weight as $\smash{\rho_N=-\rho_1-\cdots-\rho_{N-1}}$, we particularize $N-1$ roots as
\beq
\alpha_j\equiv \rho_j-\rho_N\,, \quad j=1,\dots,N-1\,,
\eeq 
which allow one to rewrite any other root as a linear combination with integer coefficients:
\beq
\alpha_{jk} & = & \rho_j-\rho_k\nonumber\\
& = & \rho_j-\rho_N+\rho_N-\rho_k=\alpha_j-\alpha_k\,,
\eeq
with $j,k\neq N$. The reason for particularizing the roots $\alpha_j$ is that it is convenient to first determine the regions delimited by the hyperplanes orthogonal to the $\alpha_j$ and then to determine how these regions are further subdivided by the hyperplanes orthogonal to the other roots.

Let us now take a point $\smash{r=r^jt^j}$ in the Cartan subalgebra.  It lies in the hyperplane orthogonal to $\alpha_j$ displaced from the origin by $2\pi \alpha_j n$ if and only if $r\cdot\alpha_j=2\pi n$ (recall that the roots are unith length vectors). To make the most of this condition, it is convenient to decompose $r$ along the basis formed by the vectors $4\pi\rho_1,\dots,4\pi\rho_{N-1}$:
\beq
r=4\pi \sum_{k=1}^{N-1}x_k\rho_k\,.\label{eq:rdecomp}
\eeq
Then,
\beq
r\cdot\alpha_j & = & 4\pi \sum_{k=1}^{N-1} x_k \rho_k\cdot\alpha_j\nonumber\\
& = & 4\pi \sum_{k=1}^{N-1} x_k\rho_k\cdot(\rho_j-\rho_N)=2\pi x_j\,,
\eeq
and, thus, the considered hyperplane corresponds to the equation $\smash{x_j=n}$. This, in turn, shows that the regions delimited by the hyperplanes associated to the $\alpha_j$ are the regions delimited by the lattice generated by vectors $4\pi\rho_1,\dots, 4\pi\rho_{N-1}$. In what follows, it will be sufficient to consider the parallelepiped defined by these vectors, corresponding to $x_k\in [0,1]$ in Eq.~(\ref{eq:rdecomp}).

Let us now see how the remaining hyperplanes further subdivide this parallelepiped. Consider an hyperplane orthogonal to $\alpha_{jk}$ and displaced by a multiple $n$ of $2\pi$ times this root. That $r$ belongs to that hyperplane means again that $r\cdot\alpha_{jk}=2\pi n$. However, we now have
\beq
r\cdot\alpha_{jk}=2\pi(x_j-x_k)\,,
\eeq
and then, the equation of the hyperplane is $x_j-x_k=n$. Since $x_j,x_k\in [0,1]$, the only hyperplanes that split the parallelepiped into non-trivial regions correspond to $n=0$, that is $\smash{x_j=x_k}$. We deduce that one possible Weyl chamber in the considered parallelepiped is the one defined by the equations $0<x_1<x_2<\cdots<x_{N-1}<1$. The other Weyl chambers in that same parallelepiped correspond to $0<x_{\sigma(1)}<x_{\sigma(2)}<\cdots<x_{\sigma(N-1)}<1$ where $\sigma$ is any permutation of $1,\dots,N-1$. This also implies that the parallelepiped is subdivided into $(N-1)!$ Weyl chambers.

These Weyl chambers can be given yet another useful characterization in terms of the defining weights. Let us consider for instance the Weyl chamber $0<x_1<x_2<\cdots<x_{N-1}<1$ and let us define the variables $\smash{y_1\equiv x_1}$, $\smash{y_2\equiv x_2-x_1}$, \dots, $\smash{y_{N-1}\equiv x_{N-1}-x_{N-2}}$ and $\smash{y_N\equiv 1-x_{N-1}}$. It is easily seen that when the $x_j$ span the considered Weyl chamber, the only constraints on the $y_k$ are $y_k\in [0,1]$ and $\sum_{k=1}^Ny_k=1$. Moreover, one can easily retrieve the $x_j$ from the $y_j$ as $\smash{x_k=y_1+\cdots+y_k}$. We can then write
\beq
\sum_{k=1}^{N-1} x_k \rho_k & = & \sum_{k=1}^{N-1} (y_1+\cdots+y_k)\rho_k\nonumber\\
& = & \sum_{k=1}^{N-1} y_k(\rho_{N-1}+\cdots+\rho_k)\,.
\eeq
It follows that the considered Weyl chamber is the convex hull of the points $\rho_{N-1}$, $\rho_{N-1}+\rho_{N-2}$, \dots, $\rho_{N-1}+\dots+\rho_1$ and $0$. Since we could have labelled the weights as wanted, the general rule is that, for any choice $\rho_1,\dots,\rho_{N-1}$ of $N-1$ weights, the convex hull of $\smash{\eta_1\equiv\rho_1}$, $\smash{\eta_2\equiv\rho_1+\rho_2}$, \dots $\smash{\eta_{N-1}\equiv\rho_1+\cdots+\rho_{N-1}}$ and $\smash{0=\rho_1+\cdots+\rho_N\equiv\eta_N}$ defines one Weyl chamber attached to the origin. The other Weyl chambers attached to the origin correspond to the convex hull of  $\smash{\eta_1\equiv\rho_{\sigma(1)}}$, $\smash{\eta_2\equiv\rho_{\sigma(1)}+\rho_{\sigma(2)}}$, \dots $\smash{\eta_{N-1}\equiv\rho_{\sigma(1)}+\cdots+\rho_{\sigma(N-1)}}$ and $0$, with $\sigma$ is any permutation of $1,\dots,N-1$.

\subsection{Confining configurations}
The previous characterization of the Weyl chambers attached to the origin is quite useful. In particular, it can be used to find symmetry invariant points within the Weyl chamber. Let us first decompose an arbitrary element of the Cartan subalgebra along the basis $4\pi\eta_1,\dots,4\pi\eta_{N-1}$:
\beq
r=4\pi\sum_{k=1}^{N-1}z_k \eta_k\,.
\eeq
Since $\eta_N=0$, we can extend this decomposition into a decomposition along an affine basis
\beq
r=4\pi\sum_{k=1}^N z_k \eta_k\,,\label{eq:affine}
\eeq
with $z_1+\cdots+z_N=1$. The considered Weyl chamber corresponds to the extra constraints $z_k\in [0,1]$. Its vertices $4\pi\eta_k$ correspond to those points with one of the coordinates $z_k$ equal to $1$ and the rest equal to $0$.

Let us now associate to each value of $r$, the ``classical'' Polyakov loop
\beq
\ell(r) & = & \frac{1}{N}{\rm tr}\,e^{ir^jt^j}=\frac{1}{N}\sum_{h=1}^N e^{ir\cdot\rho_h}\,.\label{eq:1144}
\eeq
Its value at the vertices of the Weyl chamber is
\beq
\ell(4\pi\eta_k)=\frac{1}{N}\sum_{h=1}^N e^{i4\pi\eta_k\cdot\rho_h}\,.
\eeq
Now
\beq
\eta_k\cdot\rho & = & (\rho_1+\cdots+\rho_k)\cdot\rho_h\nonumber\\
& = & \frac{\Theta(k-h)}{2}-\frac{k}{2N}\,,\label{eq:116}
\eeq
and thus
\beq
\ell(4\pi\eta_k)=\frac{1}{N}\sum_{h=1}^N e^{-i\frac{2\pi}{N}k}=e^{-i\frac{2\pi}{N}k}\,,
\eeq
where we have used that the term with the $\Theta$-function in Eq.~(\ref{eq:116}) does not contribute to the exponential in Eq.~(\ref{eq:1144}). We have thus found that the value of the classical Polyakov loop at the vertex $4\pi\eta_k$ is nothing but the $k^{\rm th}$ center element of SU(N).

Consider now a center transformation with associated center element $e^{i2\pi/N}$. Since the Polyakov loop is multiplied by this center element, we deduce that the vertices of the Weyl chamber are transformed as $4\pi\eta_1\to 4\pi\eta_2$, $4\pi\eta_2\to 4\pi\eta_3$, \dots, $4\pi\eta_{N-1}\to 4\pi\eta_N=0$ and $4\pi\eta_N=0\to 4\pi\eta_1$. From the point of view of a passive transformation, this means that the coordinates $z_k$ in Eq.~(\ref{eq:affine}) are transformed as $z_1\to z_N$, $z_2\to z_1$, \dots, $z_N\to z_{N-1}$. Then, the only invariant points are those such that $\smash{z_1=z_2=\dots=z_N}$ and, because $z_1+\cdots+z_N=1$, this common coordinate needs to be $1/N$. It follows that the center-symmetric point in the considered Weyl chamber is 
\beq
r_c & = & \frac{4\pi}{N}\sum_{k=1}^N \eta_k\nonumber\\
& = & \frac{4\pi}{N}(N\rho_1+(N-1)\rho_2+\cdots+\rho_N)\nonumber\\
& = & \frac{4\pi}{N}((N-1)\rho_1+(N-2)\rho_2+\cdots+\rho_{N-1})\,,\nonumber\\
\eeq
where we have used that $\rho_1+\cdots+\rho_N=0$.

We can similarly consider the case of charge conjugation which transforms the Polyakov loop $\ell$ associated to a particle into the Polyakov loop $\ell^*$ of the corresponding anti-particle. We deduce that, under charge conjugation, a vertex $4\pi\eta_k$ of associated Polyakov loop $e^{-i2\pi k/N}$ is transformed into the vertex of associated Polyakov loop $e^{i2\pi k/N}=e^{-i2\pi (N-k)/N}$, that is $4\pi\eta_{N-k}$. From the point of view of a passive transformation, this means that the coordinates $z_k$ in Eq.~(\ref{eq:affine}) are transformed as $z_k\to z_{N-k}$. We then need to distinguish two cases depending on whether or not there exists a $k$ such that $\smash{k=N-k}$, that is depending on whether $N$ is even or odd.

 If $N$ is odd, all coordinates are transformed into different ones: $z_1\leftrightarrow z_{N-1}$, $z_2\leftrightarrow z_{N-2}$, \dots $z_{(N-1)/2}\leftrightarrow z_{(N+1)/2}$. The charge-invariant states correspond to those elements (\ref{eq:affine}) such that $z_1=z_{N-1}$, $z_2=z_{N-2}$, \dots, $z_{(N-1)/2}=z_{(N+1)/2}$. This represents an affine space of dimension $N-1-(N-1)/2=(N-1)/2$. In the SU($3$), case this is a line, as we have recalled  above. 
 
 If $N$ is even, all coordinates are transformed into different ones, expect for $z_{N/2}$ which is mapped into itself: $z_1\leftrightarrow z_{N-1}$, $z_2\leftrightarrow z_{N-2}$, \dots $z_{N/2-1}\leftrightarrow z_{N/2+1}$, $z_{N/2}\leftrightarrow z_{N/2}$. The charge-invariant states correspond to those elements (\ref{eq:affine}) such that $z_1=z_{N-1}$, $z_2=z_{N-2}$, \dots, $z_{N/2-1}=z_{N/2+1}$, the value of $z_{N/2}$ being unconstrained. This represents an affine space of dimension $N-1-(N/2-1)=N/2$. In the SU($2$) case, this is again a line, corresponding to the whole Weyl chamber as we have already seen above. In the SU($4$) case, this would correspond to a plane, see for instance \cite{Reinosa:2020mnx}.

\subsection{Symmetry constraints}
The transformations of the Weyl chamber into itself associated to center transformations can be seen as resulting from the application of a winding transformation $V_{-\rho}(\tau)$ that translates the Weyl chamber into a different one by a vector $-4\pi\rho$, followed by a sequence of Weyl transformations which correspond to reflections with respect to the facets of the Weyl chamber, in order to bring the Weyl chamber back to its original position.\footnote{The same considerations apply to charge conjugation upon replacing the winding transformation by $A_\mu\to -A_\mu^{\rm t}$ and adapting the sequence of Weyl transformations.}

In order to find the appropriate sequence of reflections, let us consider the Weyl chamber defined by the vertices
\beq
\rho_1,\quad \rho_1+\rho_2, \quad \dots, \quad \rho_1+\rho_2+\cdots+\rho_N=0\,.\label{eq:W1}
\eeq
Under the winding transformation $V_{\rho_1}(\tau)$ it becomes the Weyl chamber of vertices
\beq
\rho_2,\quad \rho_2+\rho_3, \quad \dots, \quad \rho_2+\rho_3+\cdots+\rho_N+\rho_1=0\,.\nonumber\\\label{eq:W2}
\eeq
To continue, let us first remark that the action of the reflection w.r.t. the hyperplane orthogonal to a given root $\alpha_{jk}$ on the collections of weights $\rho_h$ is only to flip $\rho_j$ and $\rho_k$. This can be easily checked using the property (\ref{eq:theprop}).\footnote{Using this remark, one can also easily deduced that the facets of the considered Weyl chamber lie either within the hyperplanes orthogonal to $\alpha_{k(k+1)}=\rho_k-\rho_{k+1}$, with $k=1,\dots,N-1$, that go through the origin, or within the hyperplane orthogonal to $\alpha_{N1}$ displaced by $2\pi\alpha_{N1}$ with respect to the origin.} It follows that, by successively applying $W_{\alpha_{12}}$, $W_{\alpha_{23}}$, \dots $W_{\alpha_{(N-1)N}}$, one transforms the Weyl chamber (\ref{eq:W1}) into the Weyl chamber (\ref{eq:W2}). We have thus found that
\beq
U_c=W_{\alpha_{(N-1)N}}\cdots W_{\alpha_{23}}W_{\alpha_{12}}V_{-\rho_1}(\tau)\,.\label{eq:ucc}
\eeq
This can be rewritten in an alternative form using the crossing rules given in App.~C.3 and by noticing that the action of the reflection w.r.t. the hyperplane orthogonal to a given root $\alpha_{jk}$ on the collections of roots $\alpha_{j\ell}$ is to flip $\alpha_{jk}$ and $\alpha_{kj}$, $\alpha_{j\ell}$ and $\alpha_{k\ell}$, as well as $\alpha_{hk}$ and $\alpha_{hj}$. This can again be easily checked using the property (\ref{eq:theprop}). We then find
\beq
U_c=W_{\alpha_{12}}W_{\alpha_{13}}\cdots W_{\alpha_{(N-1)N}}V_{-\rho_1}(\tau)\,.
\eeq
In this form this is a generalization of Eqs.~(\ref{eq:t2}) and (\ref{eq:t3}). In the SU($3$) case, the comparison actually requires exchanging the labels $2$ and $3$ because, with the partialcular labelling (\ref{eq:rho}), the Weyl chamber that we considered in the main text is $\rho_1$, $\rho_{1}+\rho_3$, $\rho_1+\rho_3+\rho_2$. There are many other forms of $U_c$ obtained by exchanging some of the $W$'s. In what follows, we denote the general form as $U_c=W_{\alpha_{N-1}}\cdots W_{\alpha_1}V_{-\rho}(\tau)$.

The combination of the above Weyl transformations results in an isometry of the Weyl chamber into itself, centered around the confining configuration of the Weyl chamber. We note that since $V_{-\rho}(\tau)$ acts like a translation for backgrounds of the form (\ref{eq:bg}), the combined action of the Weyl  transformations only corresponds to the same isometry centered about the origin of the algebra. We denote this isometry by 
\beq
{\cal I}={\cal R}_{\alpha_{N-1}}\cdots{\cal R}_{\alpha_1}\,.\label{eq:iso}
\eeq
We now would like to analyze the constraints of the symmetry $U_c$ on the correlation functions.

Before doing so, it will be convenient to rewrite the action (\ref{eq:241})-(\ref{eq:242}) of a Weyl transformation on the various generators of the algebra in a more compact form. To this purpose, recall that another notation for $t^j$ is $t^{0^{(j)}}$ where the use of $0$ as a label emphasizes the fact that the generators $t^j$ are vanishing-charge states, while the label $(j)$ is used to distinguish these various degenerate states. Now, $0^{(j)}$ should be understood as the zero vector associated with the direction $j$ in the commuting subalgebra. We can more generally associate a zero to an arbitrary direction $u^j$. To this purpose, we define $\smash{t^{0^{(u)}}\equiv u^jt^j}$. Then, we notice that
\beq
W_\alpha t^{0^{(u)}}W^\dagger_\alpha & = & u^jt^j-2\alpha^k t^k \alpha^ju^j\nonumber\\
& = & (u^j-2(u\cdot\alpha)\alpha^j)t^j=t^{0^{(u-2(u\cdot\alpha)\alpha)}}\,.\nonumber\\
\eeq
We can now combine Eqs.~(\ref{eq:241}) and (\ref{eq:242}) into the single formula
\beq
W_\alpha t^\kappa W^\dagger_\alpha=t^{{\cal R}_\alpha\cdot\kappa}\,,\label{eq:tr}
\eeq
where $\smash{{\cal R}_\alpha\cdot\kappa\equiv\kappa-2(\kappa\cdot\alpha)\alpha}$ denotes the geometrical reflection of the vector $\kappa$ with respect to an hyperplane orthogonal to $\alpha$. In particular, under this reflection, a zero $0^{(u)}$ is mapped onto another zero $0^{(v)}$. The nuance, however, is that it is not the same zero since $\smash{v={\cal R}_\alpha\cdot u}$. Similarly, we can define the action of the isometry (\ref{eq:iso}) on any type of label $\kappa$, denoted ${\cal I}\cdot\kappa$ in what follows. It is obtained by repeated action of the ${\cal R}_{\alpha_j}$ and we note in particular that ${\cal I}\cdot 0^{(u)}\equiv 0^{({\cal I}\cdot u)}$.

Returning to the symmetry constraints associated to the transformation $U_c$, let us evaluate $U_c t^\kappa U_c^\dagger$. Upon repeated use of Eq.~(\ref{eq:tr}), it is found that
\beq
U_c t^\kappa U_c^\dagger & = & e^{-i4\pi\frac{\tau}{\beta}\rho\cdot\kappa}t^{{\cal I}\cdot\kappa}\,.
\eeq
In terms of the gauge field, this corresponds to the transformation
\beq
\delta A_\mu^\kappa\to e^{-i4\pi\frac{\tau}{\beta}\rho\cdot({\cal I}^{-1}\cdot\kappa)}\delta A^{{\cal I}^{-1}\cdot\kappa}_\mu\,.\label{eq:symm}
\eeq
Since (\ref{eq:symm}) is a symmetry within the center-symmetric Landau gauge, we obtain the following constraint on the (connected) correlation functions in this gauge:
\beq
\langle A^{\kappa_1}_{\mu_1}\cdots A^{\kappa_n}_{\mu_n}\rangle & = & e^{-i4\pi\sum_{i=1}^n\frac{\tau_i}{\beta}\rho\cdot({\cal I}^{-1}\cdot\kappa_i)}\nonumber\\
& & \times\,\langle A^{{\cal I}^{-1}\cdot\kappa_1}_{\mu_1}\cdots A^{{\cal I}^{-1}\cdot\kappa_n}_{\mu_n}\rangle
\eeq
This formula compares well with those obtained in the SU($2$) and SU($3$) case. We stress that
\beq
A_\mu^{0^{({\cal I}^{-1}\cdot u)}}={\cal I}^{-1}_{jk}u^k A_\mu^{0^{(j)}}=u^k{\cal I}_{kj} A_\mu^{0^{(j)}}\,,
\eeq
and thus, while the charged labels transform according to ${\cal I}^{-1}$, the neutral components of the field transform according to ${\cal I}$. This is also what we observe in the above examples, see for instance Eqs.~(\ref{eq:93})-(\ref{eq:95}).

\section{Conclusions}

We have performed a systematic study of the center-symmetry constraints on the correlation functions computed within the center-symmetric Landau gauge, a class of background Landau gauges where the background is chosen to be center-symmetric in a sense that we have precisely defined. We have specified to backgrounds that comply with other symmetries as well such as charge conjugation and invariance under particular color rotations, whose consequences we have also thoroughly investigated. As a result of our analysis, we have identified new signatures for the deconfinement transition from the correlation functions in those gauges, extending the results obtained in Ref.~\cite{vanEgmond:2022nuo}. 

The analysis made in that reference was restricted to the (color) neutral, chromo-electric sector because this is where the transition usually occurs. In the SU($2$) case, we found a sharp signature of the transition signalled by a divergence of the zero-momentum propagator. However, this does not really qualify as an order parameter since there exists no phase over which this quantity is constant. The present analysis shows that a more standard SU($2$) order parameter can be constructed from the (color) charged components of the propagator, both in the chromo-electric and chromo-magnetic sectors, see Eq.~(\ref{eq:z2}). Other order parameters can be constructed from the three-point function, both in the purely neutral sector (\ref{eq:z21}), or in a sector mixing neutral and charged components (\ref{eq:z22}). This can also be generalized to higher order correlators (\ref{eq:z2n}).

Similar conclusions hold for the SU($3$) gauge group. In that case, we had already identified an order parameter from the chromo-electric propagator in the neutral sector. The present analysis extends this conclusion to the chromo-magnetic sector, see Eq.~(\ref{eq:z3}), while revealing various other order parameters from the charged sector, see Eq.~(\ref{eq:z30}). The three-point function leads to four different order parameters, Eqs.~(\ref{eq:z31}), (\ref{eq:z32}), (\ref{eq:z33}) and (\ref{eq:z34}). Finally, we have extended our analysis to the SU(N) case. 

All these results confirm that the center-symmetric Landau gauge put forward in Ref.~\cite{vanEgmond:2021jyx} is a good gauge for the study of the confinement-deconfinement transition within functional approaches. Indeed, in this gauge, the transition is encoded directly within the building blocks that sustain these approaches, that is the correlation functions for the primary fields and does not require the computation of more involved order parameter such as the Polyakov loop.

In a work in preparation, we shall confront these expectations to one-loop calculations within the Curci-Ferrari model \cite{Curci:1976bt}, a model accounting for some of the low energy aspects of the gauge fixing in (background) Landau gauges \cite{Pelaez:2021tpq}. Similar calculations for the gluon three-point function are also in progress. Finally, it would be interesting to see whether similar ideas extend to the lattice implementation of center-symmetric gauge fixings. Work in this direction is also in progress.

\pagebreak

\appendix

\section{Roots and weights of SU(N)}\label{app:n}

\subsection{Weights \label{wts}}
As recalled in the main text, the (defining) weights occur when diagonalizing the defining action of the commuting generators $t^j$ of the algebra:
\beq
t^j|\rho\rangle=\rho^j|\rho\rangle\,.
\eeq
In the SU(N) case, there are $N$ weights $\rho_k$ of components
\beq\label{eq:rho}
\rho^j_k=\frac{1}{\sqrt{2j(j+1)}}\times\left\{
\begin{array}{rl}
\!1\,,\,\, & \mbox{if $k\leq j$}\\
\!-j\,,\,\, & \mbox{if $k=j+1$}\\
\!0\,,\,\, & \mbox{if $k>j+1$}
\end{array}
\right.,
\eeq
where $1\leq j\leq N-1$. It follows in particular that
\beq
\rho_k^2\equiv\rho_k\cdot\rho_k=\frac{1}{2}\left(1-\frac{1}{N}\right),
\eeq
for any $k$, and
\beq
\rho_k\cdot\rho_{k'}=-\frac{1}{2N}\,,
\eeq
for any $k$ and $k'\neq k'$. In other words
\beq
\rho_k\cdot\rho_{k'}=\frac{\delta_{kk'}}{2}-\frac{1}{2N}\,.\label{eq:theprop}
\eeq
From these properties, one can show that any strict subset of defining weights is a linearly independent set. The complete set is not a linearly independent set because it is constrained by the relation
\beq
\sum_{k=1}^N \rho_k=0\,,
\eeq
which is in fact the only constraint among the defining weights.

\subsection{Roots}
The roots occur when diagonalizing the adjoint action of the commuting generators $t^j$ of the algebra:
\beq
[t^j,t^\alpha]=\alpha^j t^\alpha\,,\label{eq:A6}
\eeq
to be added to the relations $[t^j,t^{j'}]=0$. To each root $\alpha$ is associated another root $-\alpha$. In fact, quite generally, the $t^\alpha$'s can be chosen such that $(t^\alpha)^\dagger=t^{-\alpha}$ and $[t^\alpha,t^{-\alpha}]=\alpha^j t^j$, and of course $[t^\alpha,t^\alpha]=0$. For $\beta\neq\alpha$ and $\beta\neq-\alpha$, we have instead $[t^\alpha,t^\beta]=N^{\alpha\beta}\,t^{\alpha+\beta}$, where $N^{\alpha\beta}=0$ if $\alpha+\beta$ is a not a root.

The values of $N^{\alpha\beta}$ when $\alpha+\beta$ is a root depend on the group. For SU(N), $\smash{N^{\alpha\beta}=\sigma^{\alpha\beta}/\sqrt{2}}$, where the sign $\sigma^{\alpha\beta}$ is determined in the next subsection.

\subsection{Relation between roots and weights}\label{app:rel}
The roots are not independent of the weights. In fact, they correspond to all possible differences of distinct weights. This means that any root can be written as $\smash{\alpha=\rho_\alpha-\bar\rho_\alpha}$, and it is not difficult to argue that this decomposition is unique. In particular, since if $\alpha$ is a root, $-\alpha$ is a root, we must have $\bar{\rho}_{\alpha}=\rho_{-\alpha}$. 

Also, given two roots $\alpha$ and $\beta\neq\pm\alpha$, the only possibility for $\alpha+\beta$ to be a root is that $\smash{\rho_\beta=\bar\rho_\alpha}$ or $\smash{\rho_\alpha=\bar\rho_\beta}$. This discussion is actually connected with the value of $\sigma^{\alpha\beta}$ alluded to in the previous subsection since in the first case $\smash{\sigma^{\alpha\beta}=+1}$ whereas in the second case it equals $\smash{\sigma^{\alpha\beta}=-1}$. It will be convenient to set $\sigma^{\alpha\beta}$ equal to $0$ in any other case. It can be checked that this is summarized in the following expression:
\beq
\sigma^{\alpha \beta}= 2(\rho_{\alpha}\cdot\bar\rho_{\beta}-\bar\rho_{\alpha}\cdot\rho_{\beta}), \label{sgm}
\eeq
from which we can read off the following identities:
\beq
\sigma^{\alpha \beta}&=&- \sigma^{\beta \alpha}\,,\nonumber\\
\sigma^{(-\alpha)(-\beta)}&=& - \sigma^{\alpha \beta}\,,\\
\sigma^{(-\alpha)(\alpha+\beta)}&=& \sigma^{\alpha \beta}\,.\nonumber
\eeq

From the relation between weights and roots, one can also deduce the properties (\ref{eq:31})-(\ref{eq:32}). Another important consequence which we shall exploit later on is that, for SU(N), $\alpha+n\beta$ cannot be a root if $|n|>1$.\\

\section{Preliminary calculations}\label{app:a}
Take a root $\alpha$ and a complex number $z$ and consider the transformation 
\beq
U_\alpha(z)\equiv e^{i(zt^\alpha+z^*t^{-\alpha})}\,.
\eeq 
By construction $\smash{U_{\alpha}(z)=U_{-\alpha}(z^*)}$. Moreover, using that $\smash{(t^\alpha)^\dagger=t^{-\alpha}}$, we find $U^\dagger_\alpha(z)=U_{\alpha}(-z)=U_{-\alpha}(-z^*)$. Let us now investigate how $U_\alpha(z)$ acts on the algebra.\\

\subsection{Action on the Cartan subalgebra}
Consider first the action of $U_\alpha(z)$ on a generator $t^j$ of the Cartan subalgebra. Using $e^XYe^{-X}=e^{[X,\,\,]}Y$, we can write
\beq
U_\alpha(z)\,t^j\, U^\dagger_\alpha(z) & = & e^{[i(zt^\alpha+z^*t^{-\alpha}),\,\,\,]}t^j\nonumber\\
& = & \sum_{n=0}^\infty \frac{i^n}{n!}[zt^\alpha+z^*t^{-\alpha},\,\,]^n t^j\,,\label{eq:sub}
\eeq
where $[X,\,\,\,]^nY$ stands for the nested commutator $[X,[X,[X,\dots,[X,Y]]]]$ of order $n$. The first commutators give ${[}zt^\alpha+z^*t^{-\alpha},\,\,{]}^0 t^j=t^j$ as well as
\beq
{[}zt^\alpha+z^*t^{-\alpha},\,\,\,{]}^1 t^j & = & \alpha^j(z^*t^{-\alpha}-zt^\alpha)\,,\\
{[}zt^\alpha+z^*t^{-\alpha},\,\,\,{]}^2 t^j & = & 2|z|^2\alpha^j(\alpha\cdot t)\,,
\eeq
where  we introduced the notation $\smash{\alpha\cdot t\equiv \alpha^k t^k}$. Thus, aside for the $n=0$ term in Eq.~(\ref{eq:sub}), which equals $t^j$, the series oscillates between the two operators $\smash{X_\alpha\equiv\alpha\cdot t}$ and $\smash{Y_\alpha(z)\equiv z^*t^{-\alpha}-zt^\alpha}$. It is then convenient to rewrite the above commutators as (we contract them by $\alpha^j$ and use that $\smash{\alpha^2=1}$)
\beq
{[}zt^\alpha+z^*t^{-\alpha}, X_\alpha{]} & = & Y_\alpha(z)\,,\\
{[}zt^\alpha+z^*t^{-\alpha},Y_\alpha(z){]} & = & 2|z|^2X_\alpha\,,
\eeq
from which we deduce that
\beq
{[}zt^\alpha+z^*t^{-\alpha},\,\,\,{]}^{2p+1}Y_\alpha(z) & = & (2|z|^2)^{p+1}X_\alpha\,,\label{eq:v1}\\
{[}zt^\alpha+z^*t^{-\alpha},\,\,\,{]}^{2p}Y_\alpha(z) & = & (2|z|^2)^{p}Y_\alpha(z)\,,\label{eq:v2}
\eeq
for any $p\geq 0$, or equivalently
\beq
{[}zt^\alpha+z^*t^{-\alpha},\,\,\,{]}^{2p}X_\alpha & = & (2|z|^2)^{p}X_\alpha\,,\label{eq:v3}\\
{[}zt^\alpha+z^*t^{-\alpha},\,\,\,{]}^{2p+1}X_\alpha & = & (2|z|^2)^{p}Y_\alpha(z)\,,\label{eq:v4}
\eeq
for $p\geq 0$. Using Eqs.~(\ref{eq:v1})-(\ref{eq:v2}), we find that
\begin{widetext}
\beq
U_\alpha(z)\,t^j\, U^\dagger_\alpha(z) & = & t^j+\sum_{n=1}^\infty \frac{i^n}{n!}[zt^\alpha+z^*t^{-\alpha},\,\,\,]^n t^j=t^j+\alpha^j\sum_{n=1}^\infty \frac{i^{n}}{n!}[zt^\alpha+z^*t^{-\alpha},\,\,\,]^{n-1} Y_\alpha(z)\nonumber\\
& = & t^j+\alpha^j\sum_{p=0}^\infty \frac{i^{2p+1}}{(2p+1)!}[zt^\alpha+z^*t^{-\alpha},\,\,\,]^{2p} Y_\alpha(z)+\alpha^j\sum_{p=0}^\infty \frac{i^{2p+2}}{(2p+2)!}[zt^\alpha+z^*t^{-\alpha},\,\,\,]^{2p+1} Y_\alpha(z)\nonumber\\
& = & t^j+\alpha^jY_\alpha(z)\sum_{p=0}^\infty \frac{i^{2p+1}}{(2p+1)!}(2|z|^2)^{p}+\alpha^jX_\alpha\sum_{p=0}^\infty \frac{i^{2p+2}}{(2p+2)!}(2|z|^2)^{p+1}\,.
\eeq
After resumming the series, we find eventually
\beq
U_\alpha(z)\,t^j\,U^\dagger_\alpha(z)=\,t^j+\alpha^j\Big(\cos\big(\sqrt{2}|z|\big)-1\Big)X_\alpha+i\alpha^j\frac{\sin\big(\sqrt{2}|z|\big)}{\sqrt{2}|z|}Y_\alpha(z)\,.\label{eq:res}
\eeq
\end{widetext}

\subsection{Action outside the Cartan subalgebra}
We now would like to study the action of $U_\alpha(z)$ outside the Cartan subalgebra, that is compute
\beq
& & U_\alpha(z)\,t^\beta\, U^\dagger_\alpha(z)=\sum_{n=0}^\infty \frac{i^n}{n!}[zt^\alpha+z^*t^{-\alpha},\,\,\,]^n t^\beta\,.\label{eq:ac}
\eeq
We need to consider various cases.\\

\noindent{\bf a.} 
The simplest case is when $\beta$ is equal neither to $\alpha$ nor to $-\alpha$ and, neither $\alpha+\beta$ nor $-\alpha+\beta$ are roots. Then, all nested commutators with $n\geq 1$ vanish, and we get
\beq
U_\alpha(z)\,t^\beta\, U^\dagger_\alpha(z)=t^\beta\,.\label{eq:16}
\eeq

\noindent{\bf b.} 
Next, we consider the case where $\smash{\beta=\alpha}$ or $\smash{\beta=-\alpha}$. Consider for instance $\beta=\alpha$. Then
\beq
{[}z t^\alpha+z^*t^{-\alpha},\,\,\,{]}^1t^\alpha & = & -z^*X_\alpha\,.
\eeq
Using Eq.~(\ref{eq:v3})-(\ref{eq:v4}), we then arrive at
\begin{widetext}
\beq
U_\alpha(z)\,t^\alpha\,U^\dagger_\alpha(z) & = & t^\alpha+\sum_{n=1}^\infty \frac{i^n}{n!}[zt^\alpha+z^*t^{-\alpha},\,\,\,]^n t^\alpha=t^\alpha-z^*\sum_{n=1}^\infty \frac{i^{n}}{n!}[zt^\alpha+z^*t^{-\alpha},\,\,\,]^{n-1} X_\alpha\nonumber\\
& = & t^\alpha-z^*\sum_{p=0}^\infty \frac{i^{2p+1}}{(2p+1)!}[zt^\alpha+z^*t^{-\alpha},\,\,\,]^{2p}X_\alpha-z^*\sum_{p=0}^\infty \frac{i^{2p+2}}{(2p+2)!}[zt^\alpha+z^*t^{-\alpha},\,\,\,]^{2p+1} X_\alpha\nonumber\\
& & \hspace{0.5cm}=\,t^\alpha-z^*X_\alpha\sum_{p=0}^\infty \frac{i^{2p+1}}{(2p+1)!}(2|z|^2)^{p}-z^*Y_\alpha(z)\sum_{p=0}^\infty \frac{i^{2p+2}}{(2p+2)!}(2|z|^2)^{p}\,.
\eeq
After resumming the series, we arrive eventually at
\beq
U_\alpha(z)\,t^\alpha\,U^\dagger_\alpha(z)=t^\alpha-\frac{1}{2}\left(\cos(|z|\sqrt{2})-1\right)\frac{z^*}{|z|^2}Y_\alpha(z)-i z^*\frac{\sin(|z|\sqrt{2})}{|z|\sqrt{2}}X_\alpha\,.\label{eq:27}
\eeq
To obtain the action on $t^{-\alpha}$, we use $U_\alpha(z)\,t^{-\alpha}\,U^\dagger_\alpha(z)=U_{-\alpha}(z^*)\,t^{-\alpha}\,U^\dagger_{\alpha}(z^*)$ and we arrive at
\beq
U_\alpha(z)\,t^{-\alpha}\,U^\dagger_\alpha(z)=t^{-\alpha}+\frac{1}{2}\left(\cos(|z|\sqrt{2})-1\right)\frac{z}{|z|^2}Y_\alpha(z)+iz\frac{\sin(|z|\sqrt{2})}{|z|\sqrt{2}}X_\alpha\,,\label{eq:28}
\eeq
where we have used that $\smash{X_\alpha=-X_{-\alpha}}$ and $\smash{Y_\alpha(z)=-Y_{-\alpha}(z^*)}$.
\end{widetext}

\noindent{\bf c.} 
Finally, when $\beta$ is equal neither to $\alpha$ nor to $-\alpha$ but $\alpha+\beta$ is a root, we have
we have
\beq
{[}zt^\alpha+z^*t^{-\alpha},\,\,\,{]}^1t^\beta & = & \frac{\sigma^{\alpha\beta}}{\sqrt{2}}zt^{\alpha+\beta}\,,\\
{[}zt^\alpha+z^*t^{-\alpha},\,\,\,{]}^2t^\beta & = & \frac{\sigma^{\alpha\beta}\sigma^{(-\alpha)(\alpha+\beta)}}{2}zz^*t^{\beta}=|z|^2\frac{t^\beta}{2}\,,\nonumber\\
\eeq
where we have used that $\sigma^{(-\alpha)(\alpha+\beta)}=\sigma^{\alpha\beta}$. More generally
\beq
{[}zt^\alpha+z^*t^{-\alpha},\,\,\,{]}^{2p+1}t^\beta & = & \sigma^{\alpha\beta}\frac{z|z|^{2p}}{(\sqrt{2})^{2p+1}}t^{\alpha+\beta}\,,\\
{[}zt^\alpha+z^*t^{-\alpha},\,\,\,{]}^{2p}t^\beta & = & \frac{|z|^{2p}}{(\sqrt{2})^{2p}}t^\beta\,.
\eeq
\begin{widetext}
Thus
\beq
U_\alpha(z)\,t^\beta\, U^\dagger_\alpha(z) & = & \sum_{p=0}^\infty \frac{i^{2p}}{(2p)!}[zt^\alpha+z^*t^{-\alpha},\,\,\,]^{2p} t^\beta+\sum_{p=0}^\infty \frac{i^{2p+1}}{(2p+1)!}[zt^\alpha+z^*t^{-\alpha},\,\,]^{2p+1} t^\beta\nonumber\\
& = & \sum_{p=0}^\infty \frac{i^{2p}}{(2p)!}\frac{|z|^{2p}}{(\sqrt{2})^{2p}}t^\beta+\sigma^{\alpha\beta}\sum_{p=0}^\infty \frac{i^{2p+1}}{(2p+1)!}\frac{z|z|^{2p}}{(\sqrt{2})^{2p+1}}t^{\alpha+\beta}\,.
\eeq
After resumming the series, we eventually arrive at
\beq
U_\alpha(z)\,t^\beta\, U^\dagger_\alpha(z)=\cos\left(|z|/\sqrt{2}\right)t^\beta+i\sigma^{\alpha\beta}\frac{z}{|z|}\sin\left(|z|/\sqrt{2}\right)t^{\alpha+\beta}\,,\label{eq:22}
\eeq
if $\alpha+\beta$ is a root. To obtain the corresponding formula in the case where $-\alpha+\beta$ is a root, we write $U_\alpha(z)\,t^\beta\, U^\dagger_\alpha(z)=U_{-\alpha}(z^*)\,t^\beta\, U^\dagger_{-\alpha}(z^*)$ and thus
\beq
U_\alpha(z)\,t^\beta\, U^\dagger_\alpha(z)=\cos\left(|z|/\sqrt{2}\right)t^\beta+i\sigma^{(-\alpha)\beta}\frac{z^*}{|z|}\sin\left(|z|/\sqrt{2}\right)t^{-\alpha+\beta}\,.\label{eq:24}
\eeq
\end{widetext}

\section{Weyl transformations}\label{app:b}

Weyl transformations correspond to particular cases of $U_\alpha(z)$ obtained with the choice $\smash{|z|=\pi/\sqrt{2}}$, that is $\smash{z=\pi/\sqrt{2}e^{i\theta}}$. We shall define\footnote{There is a slight abuse of notation but it should be harmless.}
\beq
U_\alpha(\theta)\equiv U_\alpha\left(\frac{\pi}{\sqrt{2}}e^{i\theta}\right)=e^{i\frac{\pi}{\sqrt{2}}(e^{i\theta}t^\alpha+e^{-i\theta}t^{-\alpha})}\,.
\eeq
By definition, we have that $\smash{U_\alpha(\theta)=U_\alpha(\theta+2\pi)}$ and $\smash{U_\alpha(-\theta)=U_{-\alpha}(\theta)}$. Moreover $\smash{U^\dagger_\alpha(\theta)=U_\alpha(\theta+\pi)}$.

\subsection{Action on the algebra}
From Eq.~(\ref{eq:res}), we find that the action of $U_\alpha(\theta)$ on the Cartan subalgebra is
\beq
U_\alpha(\theta)\,t^j\,U^\dagger_\alpha(\theta)=t^j-2\alpha^j(\alpha\cdot t)\,,\label{eq:C2}
\eeq
which does not depend on $\theta$ and corresponds to a reflection symmetry with respect to an hyperplane orthogonal to $\alpha$. The action on the rest of the algebra can be read off from Eqs.~(\ref{eq:16}), (\ref{eq:27}), (\ref{eq:28}), (\ref{eq:22}) and (\ref{eq:24}). When $\beta$ is different from $\alpha$ or $-\alpha$ and when neither $\alpha+\beta$ or $-\alpha+\beta$ are roots, we have
\beq
U_\alpha(\theta)\,t^\beta\,U^\dagger_\alpha(\theta)=t^\beta\,.\label{eq:rien}
\eeq
When $\beta$ is distinct from $\alpha$ or $-\alpha$ but $\alpha+\beta$ is a root, then
\beq
U_\alpha(\theta)\,t^\beta\, U^\dagger_\alpha(\theta)=i\sigma^{\alpha\beta}e^{i\theta}t^{\alpha+\beta}\,,
\eeq
whereas if $-\alpha+\beta$ is a root
\beq
U_\alpha(\theta)\,t^\beta\,U^\dagger_\alpha(\theta)=i\sigma^{(-\alpha)\beta}e^{-i\theta}t^{-\alpha+\beta}\,.
\eeq
Finally,
\beq
U_\alpha(\theta)\,t^\alpha\, U^\dagger_\alpha(\theta)=e^{-2i\theta}t^{-\alpha}\,,
\eeq
and
\beq
U_\alpha(\theta)\,t^{-\alpha}\, U^\dagger_\alpha(\theta)=e^{2i\theta}t^{\alpha}\,.\label{eq:deux}
\eeq

\subsection{Compact rewriting}
Interestingly, Eqs.~(\ref{eq:rien})-(\ref{eq:deux}) can be summarized into a single formula. First of all, owing to our choice of a vanishing $\sigma^{\alpha\beta}$ when $\alpha+\beta$ is not a root, as well as Eqs.~(\ref{eq:321}), we can write\footnote{The appearance of $t^{\beta-2(\beta\cdot\alpha)\alpha}$ in the RHS of Eq.~(\ref{eq:tjj}) is not a surprise. Indeed, let evaluate $[t^j,U_\alpha(\theta) t^\beta U^\dagger_\alpha(\theta)]$. Using Eqs.~(\ref{eq:canonical}) and (\ref{eq:241}), we can write
\beq
& & [t^j,U_\alpha(\theta) t^\beta U^\dagger_\alpha(\theta)]\nonumber\\
& & \hspace{0.5cm}=\,U_\alpha(\theta)[U^\dagger_\alpha(\theta) t^jU_\alpha(\theta),t^\beta ]U^\dagger_\alpha(\theta)\nonumber\\
& & \hspace{0.5cm}=\,U_\alpha(\theta)(\delta^{jk}-2\alpha^j\alpha^k)[t^k,t^\beta ]U^\dagger_\alpha(\theta)\nonumber\\
& & \hspace{0.5cm}=\,(\beta^j-2(\beta\cdot\alpha)\alpha^j)U_\alpha(\theta) t^\beta U^\dagger_\alpha(\theta)\,.\nonumber\label{eq:tjj}
\eeq
Since $U_\alpha(\theta) t^\beta U^\dagger_\alpha(\theta)$ is non-zero and does not belong to the commuting part of the algebra, we deduce first, that $\beta-2(\beta\cdot\alpha)\alpha$ is a root, and second, that
\beq
U_\alpha(\theta) t^\beta U^\dagger_\alpha(\theta)=c_{\alpha\beta}(\theta)\,t^{\beta-2(\beta\cdot\alpha)\alpha}\,.\nonumber
\eeq
}
\beq
U_\alpha(\theta)\,t^\beta\, U^\dagger_\alpha(\theta)=e^{i\frac{\pi}{2}\phi^{\alpha\beta}}e^{i\theta\psi^{\alpha\beta}}t^{\beta-2(\beta\cdot\alpha)\alpha}\,,\label{eq:114}
\eeq
where $\smash{\phi_{\alpha\beta}=\sigma^{\alpha\beta}}$ if $\alpha+\beta$ is a root, $\smash{\phi_{\alpha\beta}=\sigma^{(-\alpha)\beta}}$ if $-\alpha+\beta$ is a root, and $\smash{\phi_{\alpha\beta}=0}$ otherwise, and $\smash{\psi_{\alpha\beta}=1}$ if $\alpha+\beta$ is a root,  $\smash{\psi_{\alpha\beta}=-1}$ if $-\alpha+\beta$ is a root, $\smash{\psi_{\alpha\alpha}=-2}$, $\smash{\psi_{\alpha(-\alpha)}=2}$ and $\smash{\psi_{\alpha\beta}=0}$ otherwise.

The three possible cases for $\phi_{\alpha\beta}$ can be summarized as $\smash{\phi_{\alpha\beta}=\sigma^{\alpha \beta} +\sigma^{(-\alpha)\beta}}$, and using Eq. \eqref{sgm}, this becomes
\beq
\phi_{\alpha \beta}=-2(\rho_{\alpha}+\bar\rho_{\alpha})\cdot \beta\,.
\eeq
Similarly, it can be checked that
\beq
\psi_{\alpha \beta}=-2\alpha\cdot\beta=-2(\rho_{\alpha}-\bar\rho_{\alpha})\cdot \beta\,.
\eeq
This implies that Eq.~(\ref{eq:114}) can be rewritten as
\beq
W_\alpha t^\beta  W^\dagger_\alpha=t^{\beta-2(\beta\cdot\alpha)\alpha}\,,
\eeq
upon defining
\beq
W_\alpha\equiv U_\alpha(\theta)e^{i \pi (\rho^j_{\alpha}+\bar\rho^j_{\alpha})t^j+i2\theta(\rho^j_{\alpha}-\bar\rho^j_{\alpha})t^j}\,,\label{eq:C17}
\eeq
while the action on the $t^j$'s remains the same as before:
\beq
W_\alpha\,t^j\,W^\dagger_\alpha=t^j-2\alpha^j(\alpha\cdot t)\,.\label{eq:2299}
\eeq
We stress that, so defined, $W_\alpha$ does not depend on $\theta$. This is because, from its action on the algebra, one can reconstruct $W_\alpha$ modulo an element of the center. But the action on the algebra does not depend on $\theta$ and neither can the center-element since the center group is discrete. From the $\theta$-independence of $W_\alpha$ and Eq.~(\ref{eq:C17}), we also deduce that $U_\alpha(\theta)e^{i2\theta\alpha^jt^j}$ does not depend on $\theta$, or in other words, that the $\theta$-dependence of $U_\alpha(\theta)$ is known explicitely:
\beq
U_\alpha(\theta)=U_\alpha(\theta=0)e^{-i2\theta\alpha^jt^j}\,.\label{eq:C14}
\eeq
Another way of writing the same result is
\beq
U_\alpha(\theta_1)e^{i2\theta_1\alpha^jt^j}=U_\alpha(\theta_2)e^{i2\theta_2\alpha^jt^j}\,
\eeq
or, using $U_\alpha^\dagger(\theta)=U_\alpha(\theta-\pi)$,
\beq
U_\alpha(\theta_1)U_\alpha(\theta_2)=e^{i2(\pi+\theta_1-\theta_2)\alpha^jt^j}\,.
\eeq
In particular
\beq
U^2_\alpha(\theta)=e^{i2\pi\alpha^jt^j}\,,
\eeq
which is different from $1$ in general. On the other hand, it can be shown that $U^4_\alpha(\theta)=e^{i4\pi\alpha^jt^j}=1$. Indeed, using a basis $|\rho\rangle$ that diagonalizes the defining action of the $t^j$'s, see Eq.~(\ref{eq:defining}), we have
\beq
e^{i4\pi\alpha^jt^j}|\rho\rangle=e^{i4\pi\alpha\cdot\rho}|\rho\rangle\,,
\eeq
with $\smash{\alpha\cdot\rho\in\{-1/2,0,1/2}\}$, see Eq.~(\ref{eq:31}). 

\subsection{Crossing rules}
Now that we know that the $W_\alpha$ are a convenient way to define the Weyl transformations let us determine some useful ``crossing rules'' for the latter. To this purpose, let us evaluate $W_\alpha W_\beta W^\dagger_\alpha$. We have
\beq
W_\alpha W_\beta W^\dagger_\alpha & = & W_\alpha e^{i\frac{\pi}{\sqrt{2}}(t^\beta+t^{-\beta})}e^{i \pi (\rho^j_{\beta}+\bar\rho^j_{\beta})t^j}W^\dagger_\alpha\nonumber\\
& = & W_\alpha e^{i\frac{\pi}{\sqrt{2}}(t^\beta+t^{-\beta})}W^\dagger_\alpha W_\alpha e^{i \pi (\rho^j_{\beta}+\bar\rho^j_{\beta})t^j}W^\dagger_\alpha\nonumber\\
& = & e^{i\frac{\pi}{\sqrt{2}}W_\alpha(t^\beta+t^{-\beta})W^\dagger_\alpha } e^{i \pi (\rho^j_{\beta}+\bar\rho^j_{\beta})W_\alpha t^jW^\dagger_\alpha}\,.\nonumber\\
\eeq
Using Eqs.~(\ref{eq:241})-(\ref{eq:242}), this becomes
\beq
W_\alpha W_\beta W^\dagger_\alpha & = & e^{i\frac{\pi}{\sqrt{2}}(t^{\beta-2(\beta\cdot\alpha)\alpha}+t^{-\beta+2(\beta\cdot\alpha)\alpha})}\nonumber\\
& & \times\,e^{i \pi (\rho^j_{\beta}+\bar\rho^j_{\beta}-2(\rho_{\beta}+\bar\rho_{\beta})\cdot\alpha\alpha^j)t^j}\,,
\eeq
which starts looking like $W_{\beta-2(\beta\cdot\alpha)\alpha}$. To prove that it is indeed equal to $W_{\beta-2(\beta\cdot\alpha)\alpha}$, we need to show that
\beq
\rho_\beta-2(\rho_\beta\cdot\alpha)\alpha & = & \rho_{\beta-2(\beta\cdot\alpha)\alpha}\,,\\
\bar\rho_\beta-2(\bar\rho_\beta\cdot\alpha)\alpha & = & \bar\rho_{\beta-2(\beta\cdot\alpha)\alpha}\,.
\eeq
This seems plausible since
\beq
\beta-2(\beta\cdot\alpha)\alpha & = & \rho_\beta-\bar\rho_\beta-2(\rho_\beta-\bar\rho_\beta)\cdot\alpha\,\alpha\nonumber\\
& = & (\rho_\beta-2(\rho_\beta\cdot\alpha)\alpha)-(\bar\rho_\beta-2(\bar\rho_\beta\cdot\alpha)\alpha)\,,\nonumber\\
\eeq
but we still need to show that the two expressions between brackets in the last line correspond to defining weights. This is actually a well-known result: the set of weights is invariant under Weyl transformations. This is seen by evaluating $W_\alpha^\dagger t^jW_\alpha|\rho\rangle$. One has
\beq
W_\alpha^\dagger t^jW_\alpha|\rho\rangle & = & (t^j-2(\alpha^k t^k)\alpha^j)|\rho\rangle\nonumber\\
& = & (\rho^j-2(\rho\cdot\alpha)\alpha^j)|\rho\rangle
\eeq
and thus
\beq
t^jW_\alpha|\rho\rangle & = & (t^j-2(\alpha^k t^k)\alpha^j)|\rho\rangle\nonumber\\
& = & (\rho^j-2(\rho\cdot\alpha)\alpha^j)W_\alpha|\rho\rangle\,.
\eeq
Since $W_\alpha$ is invertible, $W_\alpha|\rho\rangle$ is non-zero and therefore $\rho-2(\rho\cdot\alpha)\alpha$ is one of the defining weights.

We have thus shown that
\beq
W_\alpha W_\beta W^\dagger_\alpha=W_{\beta-2(\beta\cdot\alpha)\alpha}\,.
\eeq
Another way of recalling this rule is in the form of crossing rules
\beq
W_\alpha W_\beta =W_{\beta-2(\beta\cdot\alpha)\alpha}W_\alpha=W_\beta W_{\alpha-2(\alpha\cdot\beta)\beta}\,.
\eeq
In words, $W_\alpha$ can cross $W_\beta$ but in doing so, it replaces $\beta$ by $\beta-2(\beta\cdot\alpha)\alpha$ in its wake.

\section{Winding transformations}\label{app:c}
Given a fundamental weight $\rho$, we define
\beq
V_\rho(\tau)=e^{i\frac{\tau}{\beta}4\pi\rho_jt^j}\,,
\eeq
known as a winding transformation. These transformations correspond to non-trivial center transformations. They act on the algebra as
\beq
V_\rho(\tau)\,t^j\,V^\dagger_\rho(\tau) & = & t^j\,,\\
V_\rho(\tau)\,t^\alpha\,V^\dagger_\rho(\tau) & = & e^{i\frac{\tau}{\beta}4\pi \rho\cdot\alpha}t^\alpha\,.
\eeq
It is useful to derive the relation between these transformations and the previously defined Weyl transformations. For instance
\beq
W_\alpha V_\rho W^\dagger_\alpha & = & W_\alpha e^{i\frac{\tau}{\beta}4\pi\rho_jt^j}W^\dagger_\alpha \nonumber\\
& = & e^{i\frac{\tau}{\beta}4\pi\rho_j W_\alpha t^jW^\dagger_\alpha} \nonumber\\
& = & e^{i\frac{\tau}{\beta}4\pi\rho_j (t^j-2\alpha^j(\alpha\cdot t))} \nonumber\\
& = & e^{i\frac{\tau}{\beta}4\pi(\rho_j-2(\rho\cdot\alpha)\alpha_j)t^j}\,,
\eeq
which rewrites more simply as
\beq
W_\alpha V_\rho W^\dagger_\alpha=V_{\rho-2(\rho\cdot\alpha) \alpha}\,.\label{eq:D5}
\eeq
Similarly
\beq
V_\rho W_\alpha V^\dagger_\rho & = & V_\rho e^{i\frac{\pi}{\sqrt{2}}(t^\alpha+t^{-\alpha})}e^{i \pi (\rho^j_{\alpha}+\bar\rho^j_{\alpha})t^j}V_\rho^\dagger\,,\nonumber\\
& = & V_\rho e^{i\frac{\pi}{\sqrt{2}}(t^\alpha+t^{-\alpha})}V_\rho^\dagger V_\rho e^{i \pi (\rho^j_{\alpha}+\bar\rho^j_{\alpha})t^j}V_\rho^\dagger\,,\nonumber\\
& = & e^{i\frac{\pi}{\sqrt{2}}V_\rho (t^\alpha+t^{-\alpha})V_\rho^\dagger}  e^{i \pi (\rho^j_{\alpha}+\bar\rho^j_{\alpha})V_\rho t^jV_\rho^\dagger}\,,\nonumber\\
& = & e^{i\frac{\pi}{\sqrt{2}}(e^{i\frac{\tau}{\beta}4\pi \rho\cdot\alpha} t^\alpha+e^{-i\frac{\tau}{\beta}4\pi \rho\cdot\alpha}t^{-\alpha})}  e^{i \pi (\rho^j_{\alpha}+\bar\rho^j_{\alpha})t^j}\,.\nonumber\\
\eeq
The first factor is nothing but $U_\alpha(\theta=4\pi\rho\cdot\alpha \tau/\beta)$. Using Eq.~(\ref{eq:C14}), we then arrive at
\beq
V_\rho W_\alpha V^\dagger_\rho=W_\alpha V_{-2(\rho\cdot\alpha)\alpha}\,.\label{eq:D7}
\eeq
The identities (\ref{eq:D5}) and (\ref{eq:D7}) can again be conveniently recast in the form of crossing rules:
\beq
W_\alpha V_\rho & = & V_{\rho-2(\rho\cdot\alpha)\alpha}W_\alpha\,,
\eeq
and
\beq
V_\rho W_\alpha & = & W_\alpha V_{\rho-2(\rho\cdot\alpha)\alpha}\,.
\eeq
Once more, $W_\alpha$ can cross $V_\rho$ but, in doing so, it replaces $\rho$ by $\rho-2(\rho\cdot\alpha)\alpha$ in its wake.

\end{document}